\pdfoutput=1	% To force pdflatex processing on arXiv
\documentclass[9pt,twocolumn,twoside]{pnas-new}
\setboolean{displaywatermark}{false}
\usepackage{natbib}
\usepackage{multicol}
\setcitestyle{square, comma, numbers,sort&compress, super}
\templatetype{pnasresearcharticle} % Choose template 
\title{Sensitivity Analysis of the Laser Power Control System to Measurement Noise in SLS 3D Printers}

\author[1]{Hamid Toshani}
\author[1]{Janith Petangoda}
\author[1]{Chatura Samarakoon}
\author[1]{Phillip Stanley-Marbell} 

\affil[1]{Physical Computation Laboratory, Department of Engineering, University of Cambridge, Cambridge CB3 0FA, UK.}
\leadauthor{Toshani} 

\keywords{Uncertainty Quantification $|$ Control Systems $|$ Additive Manufacturing.}

\usepackage{framed}
\usepackage{moreverb}
\usepackage{xspace}
\usepackage{flushend}			%

\usepackage{xfrac}			%

\usepackage{amsfonts}
\usepackage{amssymb}    
\usepackage{lipsum} % For dummy text
\usepackage{needspace} % For controlling space before new paragraphs
\usepackage[thmmarks,amsmath,hyperref]{ntheorem}
\usepackage{graphicx}
\usepackage{times}
\usepackage[scaled]{helvet} 		%
\usepackage{textcomp}
\usepackage{color}
\usepackage{colortbl}
\usepackage{booktabs}
\usepackage{listings}
\usepackage{multicol}
\usepackage{pifont}			%
\usepackage{microtype}
\usepackage{newtxmath}
\usepackage{float}
\usepackage{booktabs} % For better horizontal rules
\usepackage{siunitx} % For alignment of numbers in columns
\usepackage{cuted}
\usepackage{mmm}
\usepackage{wand-sample}
\usepackage[scaled=0.85]{beramono}
\usepackage[T1]{fontenc}
\usepackage{url}
\usepackage{multirow}
\usepackage{array}
\usepackage{subcaption}
\usepackage{geometry}
\definecolor{listinggreen}{rgb}{0,0.6,0}
\definecolor{listinggray}{rgb}{0.5,0.5,0.5}
\definecolor{listingmauve}{rgb}{0.58,0,0.82}
\definecolor{listingkeywordcolor}{rgb}{1.0,0.4,0.0}
\definecolor{listinglightgray}{rgb}{0.9863,0.9863,0.9863}

\lstdefinelanguage{FSharp}
{
	morekeywords	= {
    let,
    type,
    Measure,
	},
	sensitive	= false,
	morecomment	= [l]{\#},
	morecomment	= [s]{(*}{*)},
}

\lstdefinelanguage{Newton}
{
	morekeywords	= {
		signal,
		derivation,
		symbol,
		name,
		invariant,
		constant,
		English,
		sensor,
		name,
		none,
		dot,
		cross,
		derivative,
		integral,
		interface,
		i2c,
		spi,
		analog,
		write,
		read,
		delay,
		range,
		erasuretoken,
		uncertainty,
		accuracy,
		precision,
		Gaussian,
		exponential,
		biexponential,
		to,
		bits,
		dimensionless,
		include
	},
	sensitive	= false,
	morecomment	= [l]{\#},
	morecomment	= [s]{/*}{*/},
}

\usepackage{hyperref}
\hypersetup{
    colorlinks=false, %
    linktoc=all,     %
    linkcolor=blue,  %
}

\DeclareMathOperator*{\argmin}{argmin}
\begin{abstract}
Uniform temperature distribution in Selective Laser Sintering (SLS) is essential for producing durable 3D prints. Achieving uniformity requires a laser power control system that minimises deviation of the printing temperatures from the target temperature. Because the estimate of the actual process temperature is an input to the laser power control, uncertainty in the estimate of the actual temperature can lead to fluctuations in laser power that affect the thermal performance of the SLS. This article investigates the sensitivity of a laser power control system to temperature measurement uncertainty. This article evaluates the effectiveness of two methods for quantifying the effect of input uncertainty on a SLS laser power control system: a recent innovation in uncertainty-tracked architecture and traditional Monte Carlo simulation. We show that recent advances in computer architecture for arithmatic on probability distributions make it possible for the first time, to perform control system uncertainty analysis with latencies under 30 ms, while achieving the same level of uncertainty analysis as Monte Carlo methods with latencies that are two orders of magnitude slower. \end{abstract}

\begin{document}

% Optional adjustment to line up main text (after abstract) of first page with line numbers, when using both lineno and twocolumn options.
% You should only change this length when you've finalised the article contents.
\verticaladjustment{-5pt}

\maketitle
\thispagestyle{firststyle}
\ifthenelse{\boolean{shortarticle}}{\ifthenelse{\boolean{singlecolumn}}{\abscontentformatted}{\abscontent}}{}

% If your first paragraph (i.e. with the \dropcap) contains a list environment (quote, quotation, theorem, definition, enumerate, itemize...), the line after the list may have some extra indentation. If this is the case, add \parshape=0 to the end of the list environment.
% \section{Introduction}
% \label{section:introduction}
\vspace{-0.1in}
\dropcap{S}{elective} Laser Sintering (SLS)
3D printing of materials such as PA12 nylon powders offers an efficient approach to manufacturing specialised and intricate product designs~\cite{cai2021comparative, awad20203d}. Using a layer-by-layer process based on a 3D CAD model, this technology enables the creation of structures that may be unattainable by conventional manufacturing methods~\cite{han2022advances}. 

\begin{figure}[!t]
    \centering
        \includegraphics[width=0.95\linewidth, trim={0cm 0cm 0cm 0cm}, clip]{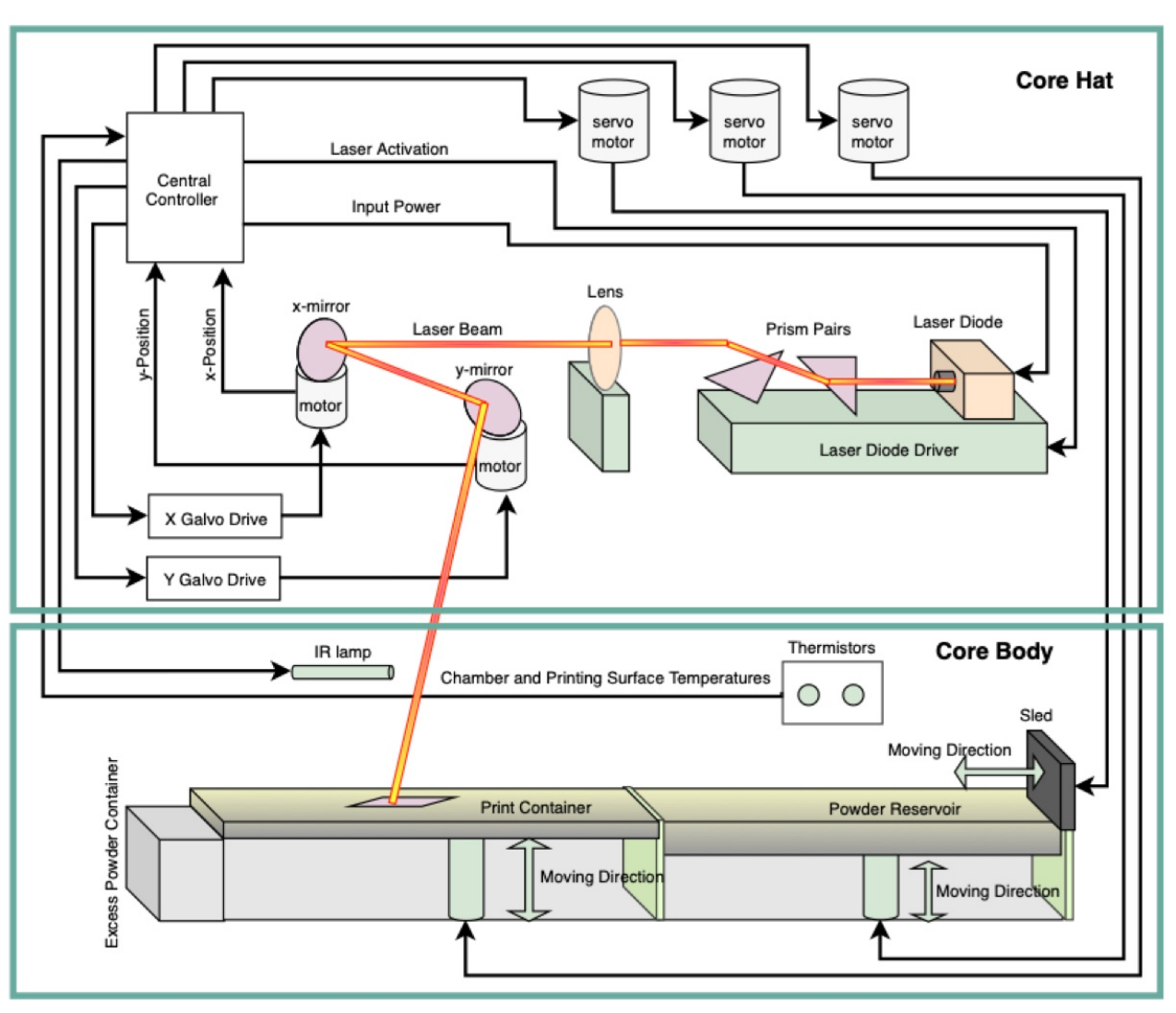}
    \caption{Illustration depicting the main components and signal flow of the SLS 3D printer. The printer moves the laser beam generated by a laser diode system across the printing bed to track the laser scanning path during the sintering process.}~\label{fig:sls-diagram}
\end{figure}

Figure \ref{fig:sls-diagram} illustrates the connections between the main components of a SLS 3D printer. Two mirrors direct the laser beam to track a pre-determined scanning path. Two galvanometers change the mirror directions. The laser beam traverses the printing bed at a pre-defined speed. After sintering a layer, the build platform moves downwards while the reservoir platform moves upwards. A roller evenly distributes the powder for sintering the next layer. This process continues layer by layer until the 3D object is fully fabricated~\cite{thakar20223d}. 

To ensure the production of printed objects with desirable mechanical properties such as strength and surface smoothness, consistent temperature distribution across the printing bed is critical. Extreme horizontal and vertical temperature variations can lead to undesirable results such as warping of the component during the manufacturing process or sub-optimal mechanical and dimensional properties~\cite{benda1994temperature, jabri2022review}. The ultimate tensile strength of an SLS component also depends on the temperature reached during the sintering process~\cite{wroe2016situ}. 

Exploring the thermal distribution of laser spots during the sintering process offers several benefits. Thermal analysis in laser-sintering processes allows for the anticipation of thermal stresses and microstructures in fabricated parts and ensures that polymer degradation does not compromise component properties~\cite{vasquez2011optimum, desynchrotron}. It aids in the creation of components possessing the intended mechanical characteristics and dimensional accuracy~\cite{zeng2012review, ibraheem2002thermal}. Temperature field simulation contributes significantly to understanding the impacts of the laser power and spot size on the temperature distribution during sintering~\cite{he2018temperature, ren2011simulation, ma2007temperature}. 
\begin{figure}[t]
    \centering
        \hspace*{-0.6cm}
        \includegraphics[width=1\linewidth, trim={0cm 0cm 0cm 0cm}, clip]{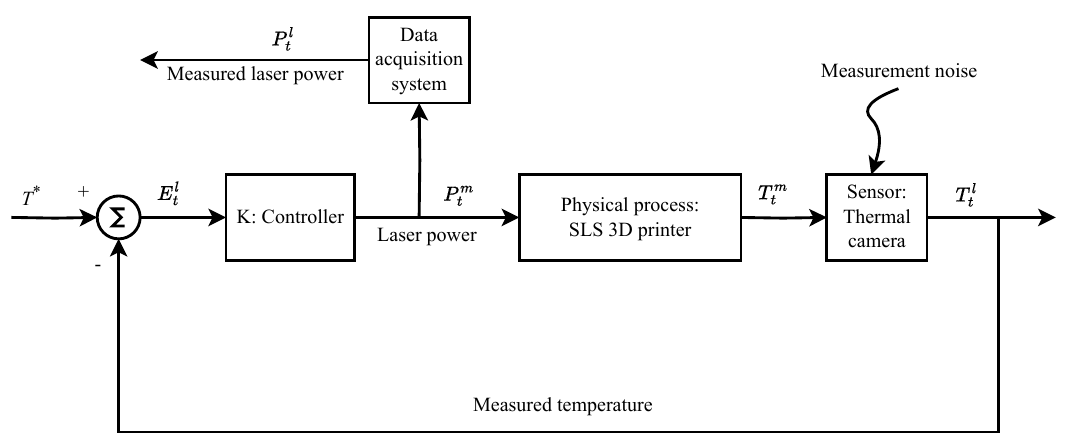}
        \caption{Block diagram of a laser power control system in an SLS 3D printer. The controller generates the laser power $P_t^{m}$ to minimise the tracking error $E_t^{l}$, which is the difference between the desired temperature $T^{*}$ and the measured temperature $T_t^{l}$. The measured temperature is the sum of the output temperature $T_t^{m}$ and a random value of the measurement uncertainty. The primary objective of the controller is to ensure a consistent temperature distribution during the sintering. Temperature measured by the thermal camera can be affected by uncertainty arising from measurement noise and sensor quantization errors. This uncertainty is subsequently transmitted through the controller, leading to fluctuations in laser power. As a result, it can substantially alter the thermal distribution throughout the sintering process.}~\label{fig:laser-power-control}
\end{figure}

The main sources of thermal variation in the SLS process are deficiencies in the heating system, laser power, and scan patterns that result in actual temperature that differs from the desired temperature.~\cite{zhang2018automatic, kiani2020investigation, lopes2022influence}. A laser power control system aims to adjust the laser power to minimise the error between the desired temperature and the temperature of the sintering point. 

Figure \ref*{fig:laser-power-control} shows the block diagram of a laser power control system. Let $P_t^l$ and $T_t^l$ be the measured laser power and the measured temperature at each time $t$. Then, we define the tracking error $E_t^l$ as the difference between $T_t^l$ and desired temperature $T^*$. 

The controller dynamically changes the laser power to ensure process stability when combined with a controller, minimises steady-state tracking error, reduces sensitivity to uncertainties in either the measurements or the parameters of the physical process, optimises energy consumption, and minimises overshoot during transitions from transient to steady-state conditions~\cite{doyle2013feedback}. 

This article focuses on the optimised control system, assuming there are optimal parameters $({k_1^*, k_2^*, ..., k_r^*})$ of the controller $K(k_1,k_2,...,k_r,t$), which leads to the minimum steady-state tracking error $E_{ss}$. That is,

\begin{equation}\label{temp}
    {k_1}^*, {k_2}^*, ..., {k_r}^* = \argmin_{k_1, k_2, ..., k_r} E_{ss}(k_1,k_2,...,k_r)
\end{equation}
The relationship between $E_t^l$ and $E_{ss}$ can be expressed as $E_{ss} = \lim_{t \to \infty} E_t^l$. 

Typically, we optimise control parameters under the assumption that the measurements are free of uncertainty (nominal conditions~\cite{mahmoud2012nominal}). However, real control systems operate in an environment where measurement uncertainties exist. 

For example, in a laser power control system, temperature uncertainties arise from sources such as measurement noise or sensor quantisation errors. These uncertainties lead to unwanted fluctuations in tracking error and laser power, which ultimately result in uneven temperature distribution during the sintering process. This article therefore investigates the impact of measurement uncertainty on control systems designed for SLS additive manufacturing. We investigate both epistemic as well as aleatoric uncertainty by injecting uniform and Gaussian additive temperature measurement uncertainties, respectively. We examine the influence of these uncertainties on the steady-state tracking error of the laser power control system without taking uncertainty into account.

To assess the impact of uncertainty on control systems, we first derive $E_{ss}$ without measurement uncertainty, which we call $e_{ss}$. We choose control parameters that minimise $e_{ss}$; we call these parameters the nominal control parameters. Keeping the nominal control parameters fixed, we then introduce temperature measurement uncertainty into the system. The introduction of measurement uncertainty makes the steady-state tracking error a random variable $E_{ss}$. We evaluate the variability of the control system performance under temperature measurement uncertainty as the variance of $E_{ss}$. 

To obtain the probability distribution of the random variable $E_{ss}$, we use two methods. The first method is the Monte Carlo simulation~\cite{zio2013monte}. Here we generate $m$ samples of $E_{ss}$ by simulating the control system with $n$ control iterations, sampling the measurement noise $m$ times at each time step. This means that at each control iteration of the control system, we take $m$ sample of the measurement noise from its probability distribution and apply it to the true measurement. We do this for $m$ time steps, taking the final measured steady-state error $e_m$ as a sample of $e_{ss}$. 

The second method is to use Laplace, which is a recently developed computer architecture that intrinsically handles uncertainty and computations involving uncertain values~\cite{tsoutsouras2022laplace}. We represent the temperature measurement uncertainty as a probability distribution, which is represented in terms of Laplace's internal representation for probability distributions. Thus, we obtain the distribution of $E_{ss}$ by running the simulation of the control system once for $n$ time steps, without carrying out the $m$ repetitions we did for the Monte Carlo simulation.

\vspace{5pt}
\noindent\textbf{Contributions}\newline
\vspace{2pt}
This article makes the following contributions to the state of the art:
\newcommand*\circled[1]{\kern-1.5em%
  \put(0,3){\color{black}\circle*{8}}\put(0,3){\circle{8}}%
  \put(-2,0.5){\color{white}\bfseries\small#1}~~}

\begin{enumerate}[label=\protect\circled{\arabic*}]
    \item \textbf{Methodology for analysing sensitivity to measurement uncertainty}. We present a framework for statistically evaluating the sensitivity of a nominally designed laser power control system to steady-state tracking error. We show that temperature measurement uncertainty can significantly affect the performance of a nominally designed laser power control system, resulting in deviations in tracking error between -2.5~\textdegree C and 2.5~\textdegree C compared to the nominal error. It is therefore crucial to take measurement uncertainty into account when designing the controller for the laser power control system of the SLS 3D printer.
            
    \item \textbf{Alternative approach to Monte Carlo simulation for sensitivity analysis}. We first use Monte Carlo simulation to calculate the steady-state error distribution. We then use an uncertainty-tracked computer architecture to compute the error distribution. We show that for the same level of accuracy, the uncertainty-tracked computer architecture can perform the analysis 17$\times$ to 71$\times$ faster than traditional Monte Carlo simulation under Gaussian and uniform temperature measurement uncertainties, respectively. These speedups make it plausible that sensitivity analyses such as those we present could in the future be performed on-line and in real-time.
\end{enumerate}

\section{Preliminaries}
Let $T_t^m$ be the printing surface temperature and $\varepsilon$ be the uncertainty in temperature measured by a thermal camera. We represent this uncertainty using two different types of distributions: a Gaussian distribution $\varepsilon\sim N(0 , \sigma^2)$ with mean zero and variance $\sigma^2$, and a uniform distribution $\varepsilon\sim U(u_{min},u_{max})$ with lower and upper bounds $u_{min}$ and $u_{max}$. Let, $\varepsilon_t$ be a random sample of the $\varepsilon$. We then describe the measured temperature $T_t^l$ as
\begin{equation}\label{measured_temp}
    T_t^l = T_t^m + \varepsilon_t
\end{equation}
Since the controller uses $T_t^l$ to update the laser power, measurement uncertainty $\varepsilon_t$ initially impacts the tracking error $E_t^l$ and subsequently influences laser power $P_t^l$ by passing through the controller $K$. 
We aim to demonstrate how measurement uncertainty in temperatures causes deviations in tracking error and laser power from their nominal values. 

Because of uncertainty in $T_t^l$, both tracking error and laser power are random variables, which means that they have a probability distribution at every time step. To assess uncertainty in these signals, we employ Monte Carlo simulation. Subsequently, we compare the performance of traditional Monte Carlo simulation with a recently-developed alternative to Monte Carlo that promises much better analysis speeds.

The Wasserstein distance is a metric quantify the distance between two probability distributions. Let P and Q be two probability distributions over a metric space (M, d), where d is a distance on M. 

The p-Wasserstein distance between P and Q probability distributions is~\cite{peyre2019computational}
\begin{equation}\label{Wasserstein}
    W_p(P, Q) = \left(\inf_{\pi} \int d(x, y)^p ,d\pi(x, y)\right)^{1/p}
\end{equation}
where $\pi$ ranges over all joint distributions whose marginals are P and Q respectively.
We assess how far the generated outputs deviate from the true answer by calculating the Wasserstein gap between the output distribution from each approach and the actual true output distribution. To determine the true output distribution, we run the Monte Carlo method with 1,000,000 number of Monte Carlo iterations.
% \vspace{1em} 
\subsection{Uncertainty in calibrated temperature measurement}
\indent\vspace{2pt}

In this section, we investigate the effects of uncertainties in the calibration parameters on the calibrated temperature measurements. As a case study, we examine the mathematical equation used by an FLIR thermal camera to convert uncalibrated radiometric measurements into calibrated temperatures.

\begin{figure}[h]
    \centering
    \includegraphics[ clip=true, width=0.47\textwidth]{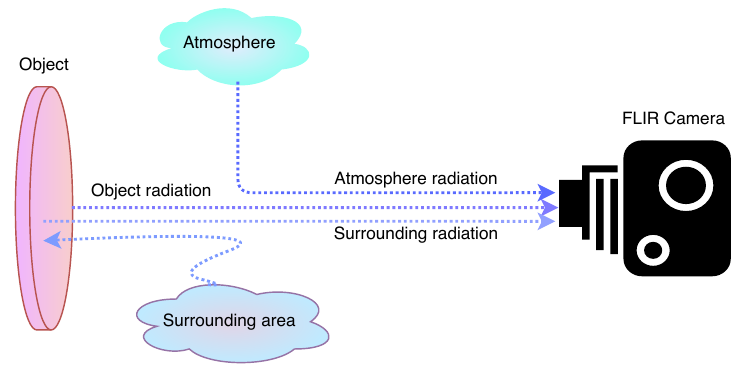}
    \vspace{0pt}
    \caption{A thermographic measurement using an FLIR camera captures radiation emitted by the object, as well as radiation reflected from the object's surface originating from the surroundings, and radiation passing through the atmosphere.}~\label{fig:flir}
\end{figure}

Let \(W_{obj}\) be the incident radiation power from the object, \(W_{refl}\) be the incident radiation power from the surroundings reflected by the object's surface, \(W_{atm}\) be the incident radiation power from the atmosphere, \(\epsilon\) be the emissivity, and \(\tau\) be the transmittance. Figure~\ref*{fig:flir} illustrates a general thermographic measurement scenario between an FLIR camera and an object. In addition to the radiation emitted by the object (\(\epsilon \tau W_{obj}\)), the camera also receives radiation from the atmosphere \(((1 - \tau) W_{atm})\) and radiation from the surroundings reflected by the object's surface \(((1 - \epsilon) \tau W_{refl})\)~\cite{flir_temperature_formula}.

\vspace{-10pt} 
Let $d$ be the distance between the environment and the object surface, $T_{refl}$ the ambient temperature, $T_{AtmC}$ the temperature of the atmosphere in degrees Celsius, $T_{Atm}$ the temperature of the atmosphere in Kelvin, $H$ the humidity of the atmosphere, $T_E$ the temperature of the external optics and $\tau_E$ the transmittance of the external optics. Let $R$, $B$, $F$, $J_0$ and $J_1$ be the camera calibration parameters. The equation that converts the uncalibrated radiometric measurements ($x_{raw}$) into calibrated temperatures ($T_{calib}$) is given by equation (4)~\cite{flir_temperature_formula}.
\begin{figure}[!t]
    \centering
    \includegraphics[ clip=true, width=0.47\textwidth]{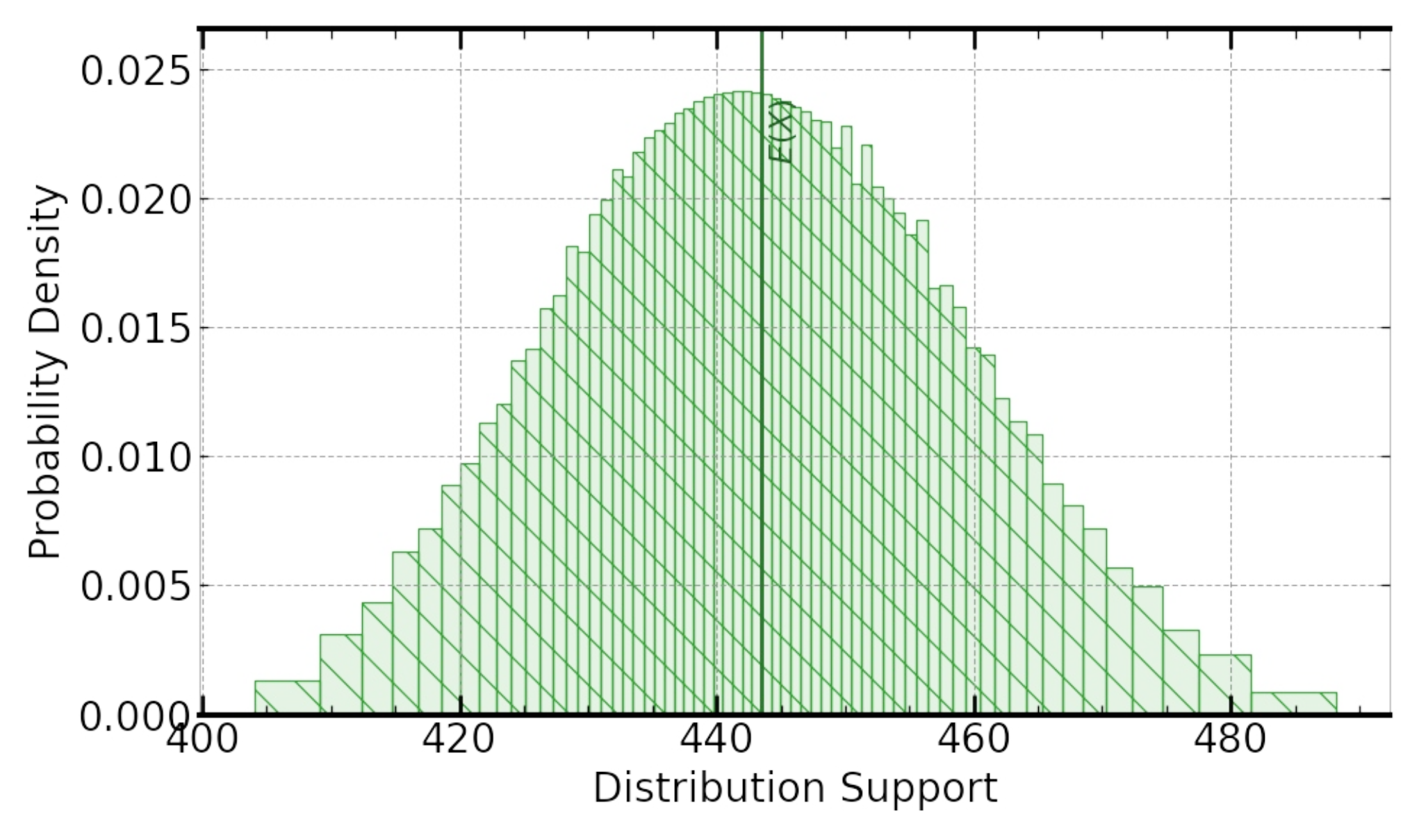}
    \vspace{0pt}
    \caption{Distribution of the calibrated temperature in the presence of uncertainty in the calibrated parameters. The uncertainty in the calibration parameters alone leads to an uncertainty in the calibrated temperature distributed between 400\,K and 480\,K.}~\label{fig:flir_uncertainty}
\end{figure}

\begin{figure*}[t]
    \centering
    \begin{equation}
    T_{calib} = \frac{B}{\ln \left( F + \frac{R}{\frac{(data_{raw} - J_0)}{J_1 \cdot \tau \cdot \epsilon \cdot \tau_E} - \left[ \frac{(1 - \epsilon) \cdot R}{\epsilon \cdot \exp \left( \frac{B}{T_{refl}} \right) - F} + \frac{(1 - \tau) \cdot R}{\epsilon \cdot \tau \cdot \left( \exp \left( \frac{B}{T_{Atm}} \right) - F \right)} + \frac{(1 - \tau_E) \cdot R}{\epsilon \cdot \tau \cdot \tau_E \cdot \left( \exp \left( \frac{B}{T_E} \right) - F \right)} \right] } \right)} - 273.15
    \end{equation}
    \label{eq:calibration}
\end{figure*}

For each calibration parameter given by FLIR, we treat them as a uniform distribution to represent the epistemic uncertainty of that calibration constant. FLIR's calibration constant 'B' becomes an input uniform distribution $B\sim \mathcal{U}(1428.0 - 0.05, 1428.0 +0.05)$. We represent other calibration parameters as $R\sim \mathcal{U}(16556.0 - 0.05, 16556.0 +0.05)$, $F\sim \mathcal{U}(1.0 - 0.05, 1.0 +0.05)$, $J_1\sim \mathcal{U}(22.5916 - 0.00005, 22.5916 +0.00005)$, $J_0\sim \mathcal{U}(89.796 - 0.0005, 89.796 +0.0005)$, $\epsilon_E\sim \mathcal{U}(1.0 - 0.05, 1.0 +0.05)$, $T_E\sim \mathcal{U}(20.0 - 0.05, 20.0 +0.05)$, $\tau\sim \mathcal{U}(1.0 - 0.05, 1.0 +0.05)$, $h\sim \mathcal{U}(0.0/100 - 0.05/100, 0.0/100 + 0.5/100)$, $T_{Atmc}\sim \mathcal{U}(21.85 - 0.005, 21.85 +0.005)$, $T_{refl}\sim \mathcal{U}(0.0 - 0.05, 0.0 + 0.05)$, $d\sim \mathcal{U}(16556.0 - 0.05, 16556.0 +0.05)$, $\epsilon\sim \mathcal{U}(1.0 - 0.05, 1.0 +0.05)$, $T_{Atm}\sim \mathcal{U}(295.0 - 0.005, 295.0 +0.005)$. 

We treat all calibration parameters as distributions based on the implied uncertainty in the significant digits. We then run a Monte Carlo simulation to estimate the calibrated temperature output for a given uncalibrated radiometric measurement. 

Figure~\ref*{fig:flir_uncertainty} shows the distribution of a calibrated temperature derived from the equation (4) associated with a raw thermal camera reading of 59000. In the absence of uncertainty in both the raw camera reading and the calibration parameters, the calibrated temperature is 442.79\,K. When the uncertainty in the calibration parameters is modeled as uniform distributions, the calibrated temperature follows a Gaussian distribution, ranging from 400\,K to 480\,K. 

This variation indicates an uncertainty of approximately ±40\,K, reflecting the significant impact of uncertainty in the calibration parameters on the calibrated temperatures, even when the raw camera readings are assumed to be precise.

\section{Laser power control system} 
\label{section:laser-power-control}
Thermal modelling using heat transfer and heat source models is a critical task in creating a laser power control system that adjusts laser power based on laser spot temperatures. Let $\rho$ be the material density ($\text{kg/m}^3$), $c$ be the specific heat capacity ($\text{J/kg}\cdot\text{K}$), $k$ be the thermal conductivity ($\text{W/m}\cdot\text{K}$), $T_t^m$ be the temperature and $Q_t^m$ be the volumetric heat generation. The heat transfer model in $x$, $y$ and $z$ directions is~\cite{hussein2013finite}:
\begin{align}\label{heat-transfer}
    \rho c\frac{\partial T_t^m}{\partial t} = \frac{\partial}{\partial x}(k\frac{\partial T_t^m}{\partial x}) + \frac{\partial}{\partial y}(k\frac{\partial T_t^m}{\partial y}) + \frac{\partial}{\partial z}(k\frac{\partial T_t^m}{\partial z}) + Q_t^m
\end{align}
The Gaussian distribution frequently emerges as the predominant beam profile in laser material processing. Let $(x_0,y_0,z_0)$ be the coordinates of the instantaneous centre of the laser beam at the print bed, $A$ the absorptivity of the powder material, $w$ the radius of the beam, $d_p$ the penetration depth and $P_t^m$ the laser power. The volumetric heat source is~\cite{hussein2013finite}
\begin{align}\label{heat-source}
    \small
    \displaystyle Q_t^m = \frac{2AP_t^m}{\pi w^2d_p}\exp(-2[ \frac{({x - x_0}^2 + {x - x_0}^2)}{w^2} + \frac{{z - z_0}^2}{d_p^2}])
 \end{align}
 The main purpose of a laser power control system is to determine the laser power to adjust the temperature of the printing bed at a desired temperature of $T^*$. We use a Proportional-Integral-Derivate (PID) controller that uses only the error $E_t^l$ between the measured temperature $T_t^l$ and setpoint $T^*$. A PID has a proportional term with gain $K_p$, an integral term with gain $K_i$, and a derivative term with gain $K_d$~\cite{aastrom2006advanced}:
 \begin{align}\label{pid}
    P_t^m= K_pE_t^l + K_i \int_{0}^{t} E_t^l \,dt + K_d\frac{dE_t^l}{dt}  
 \end{align}
 To adjust the printing bed temperature at 167.5~\textdegree C, we tune the PID parameters as $K_p=0.1$, $K_i = 0.05$, and $K_d = 5\times 10^{-5}$.
 
\section{Sensitivity analysis}
\label{section:Sensitivity-analysis}
In a laser power control system, the tracking error, which is the difference between the temperature measured by the thermal camera and the desired temperature, serves as the basis for updating the control signal. This update results in an adjusted laser power for the sintering process. The accuracy of temperature measurements is subject to uncertainties due to quantization errors in the thermal camera and background noise. These uncertainties directly affect the tracking error, leading to variations in the error, which in turn affect the behaviour of the laser power and consequently the overall sintering performance.

This section focuses on analysing the sensitivity of a laser power system to measurement uncertainties in the temperatures recorded by a thermal camera. To perform the sensitivity analysis, we use Monte Carlo simulation as a traditional approach to calculate the tracking error and laser power distributions. We then use Laplace, a new advanced computer architecture, to quantify the uncertainty in the tracking error and compare its results with the Monte Carlo simulation in terms of Wasserstein distance, a measure of how close the distributions of two methods are to a ground truth distribution, and required run time.
\subsection{Temperature measurement uncertainty in the laser power control system}
\indent\vspace{2pt}

We assume that the temperature sensor's output could have either Gaussian or uniform measurement uncertainty. To determine the possible range of this measurement uncertainty, we first explored the accuracy of the MLX90640 IR thermal camera that we used to capture temperatures during the sintering process. This sensor typically has an accuracy of plus or minus 3\,\textdegree C for temperatures between -40\,\textdegree C and 300\,\textdegree C. We also measured the printing temperature through an experimental setup that included the MLX90640 IR thermal camera and an FPGA board.
\begin{figure}[!b]
    \centering
    \includegraphics[ clip=true, width=0.47\textwidth]{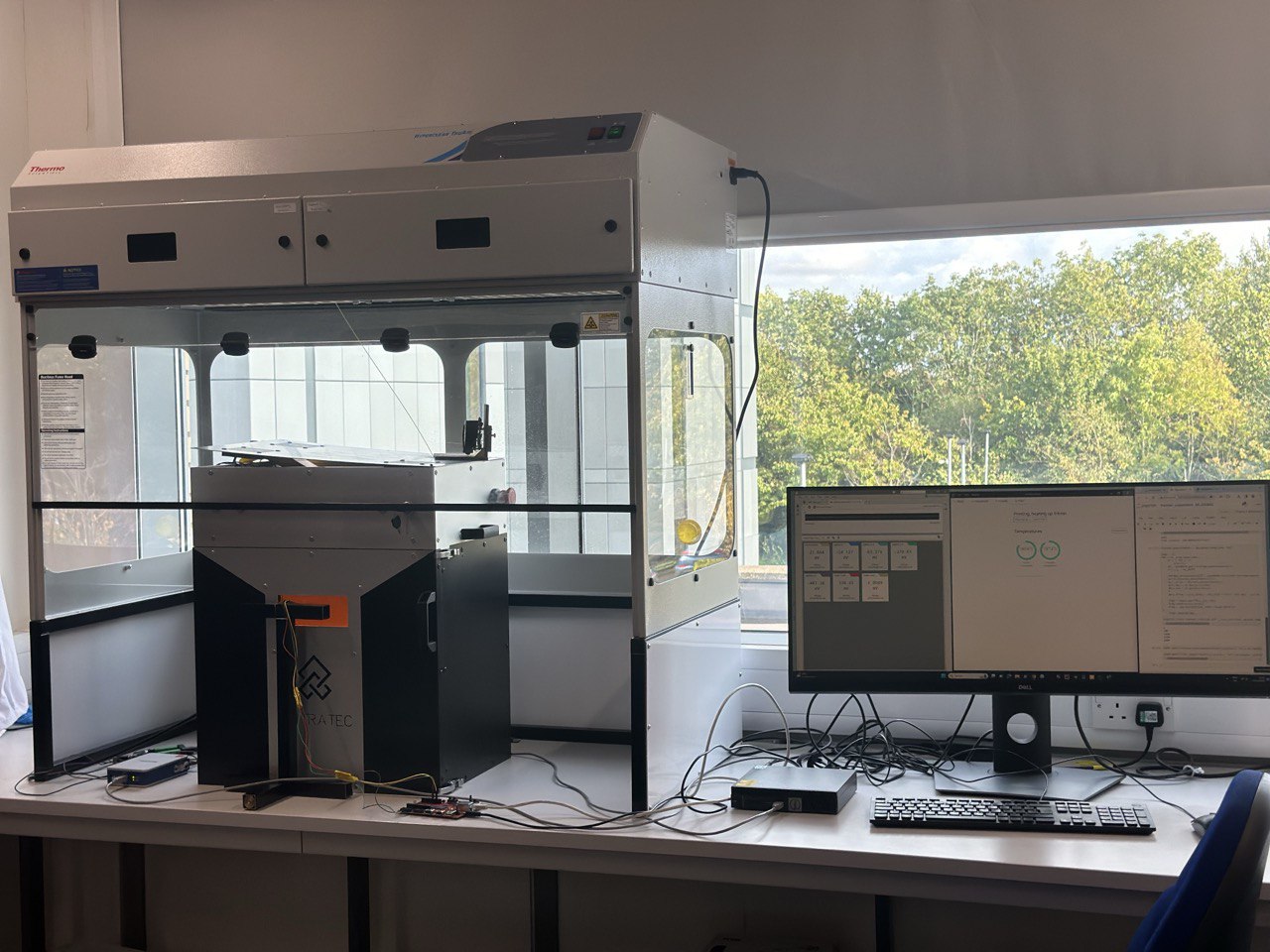}
    \vspace{0pt}
    \caption{Experimental setup for measuring the temperatures of the printing bed during the sintering process. We use MLX90640 IR thermal camera with field of view 24$\times$32 integrated into an FPGA board.}~\label{fig:measurement_implement}
\end{figure}

\begin{figure}[!t]
    \centering
    \begin{subfigure}{0.5\linewidth}
        \centering
        \includegraphics[width=\linewidth]{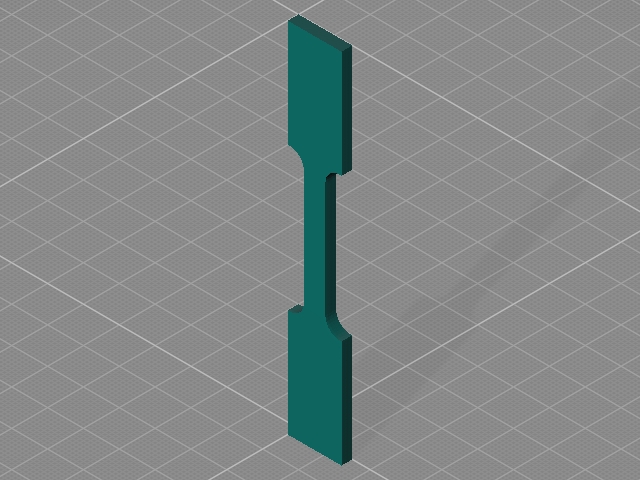}
        \caption{3D tensile specimen}
    \end{subfigure}%
    \begin{subfigure}{0.4\linewidth}
        \centering
        \includegraphics[width=\linewidth]{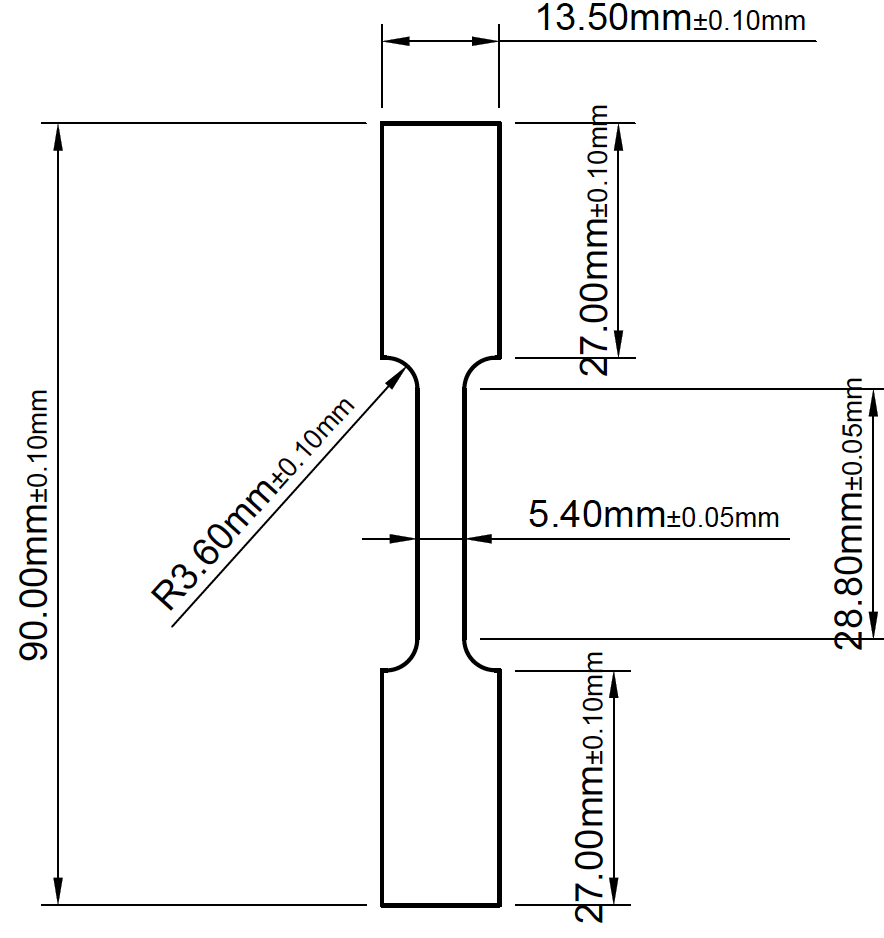}
        \caption{Tensile dimension}
    \end{subfigure}
    \caption{We print objects in the shape of a tensile specimen. This familiar geometry is widely used in tensile testing and serves as a well-established approach to assessing attributes such as material strength, ductility and various mechanical properties when subjected to tension. The dimensions of the object fit comfortably within the recommended print area for the SLS printer, which is 90 mm $\times$ 90 mm.}\label{fig:tensile}
\end{figure}

Figure~\ref*{fig:measurement_implement} shows the experimental setup for real-time measurement of the surface temperature during sintering. We collected temperatures while printing tensile specimen objects. These objects commonly have a standardized cross-section design, consisting of two shoulders and an intervening gauge. Figure~\ref*{fig:tensile} illustrates the 3D geometry of the tensile specimen object. Its total length is 90mm, while its grip cross-section spans 27mm by 13.5mm. The gauge length extends to 28.8mm, featuring a gauge diameter of 5.4mm, and there is a 25mm gap between the shoulders. 

\vspace{-2pt}
\begin{figure}[!b]
    \centering
    \includegraphics[clip=true, width=0.485\textwidth]{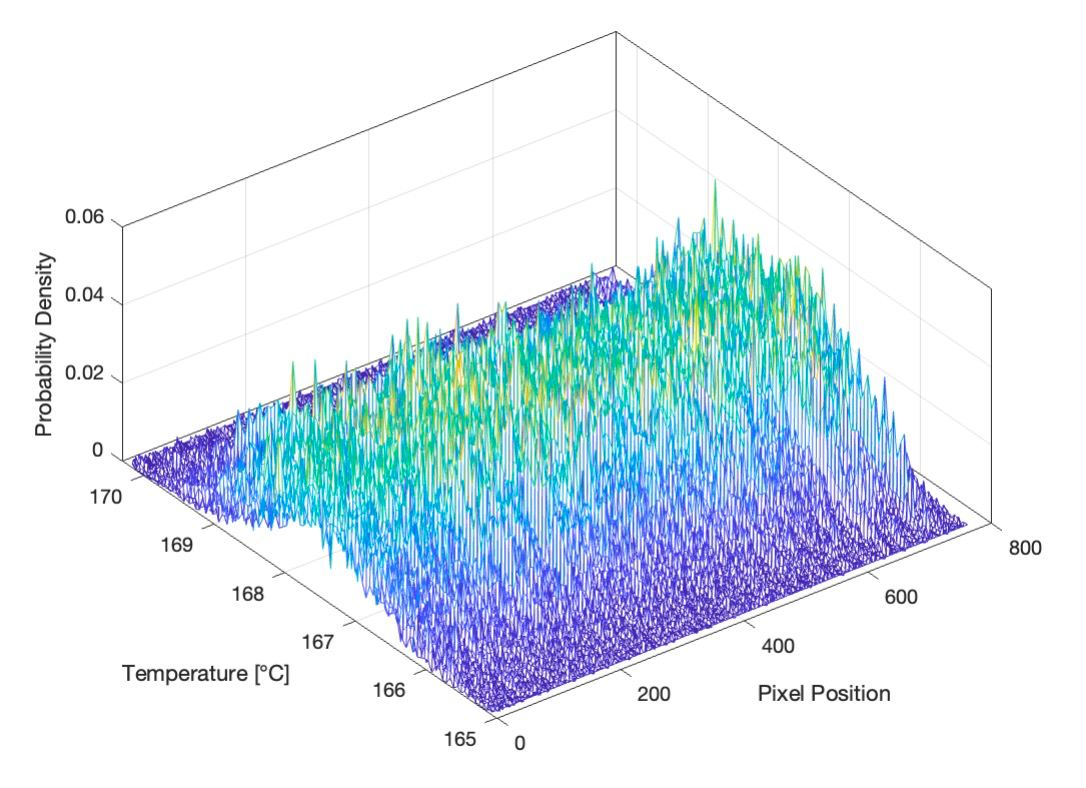}
    \caption{Temperature measurements taken by the MLX90640 IR thermal camera during the sintering process. The data contains 2125 temperature readings, with each reading corresponding to one of the 768 pixel positions on the camera. Assuming the desired temperature is 167.5\,\textdegree C, the possible range of measurement error is -3\,\textdegree C to 3\,\textdegree C due to the accuracy limitations of the camera.}~\label{fig:mlx_temp}
\end{figure}

We set the layer height, hatching distance and scan speed to 100 microns, 175 microns, and 450 mm/sec respectively. Figure~\ref*{fig:mlx_temp} shows the distribution of temperatures measured with the MLX90640 IR thermal camera. The results show that the possible range of temperature variations for each pixel position is between 164.6\,\textdegree C and 170.5\,\textdegree C. 

We assume that the desired temperature is 167.5\,\textdegree C, which means that the possible range of measurement uncertainty is approximately from -3\,\textdegree C to +3\,\textdegree C. We use this range to set the parameters for uniform and Gaussian temperature measurement uncertainties in our simulations.

\subsection{Distributional description of tracking error}
\indent\vspace{2pt}

In practice, there is uncertainty in calibration parameters for converting raw sensor ADC readings into calibrated values. As a result, even in the absence of uncertainty in measurements, calibrated temperature readings for the SLS laser power control system will have uncertainties. This implies that during each control iteration, we should describe a measured temperature as a distribution representing uncertainty in the measured temperature, rather than a single value. 

\begin{figure}[!t]
    \centering
    \begin{subfigure}{0.9\linewidth}
        \centering
        \hspace*{-0.6cm}
        \includegraphics[width=1.1\linewidth, trim={0cm 0cm 0cm 0cm}, clip]{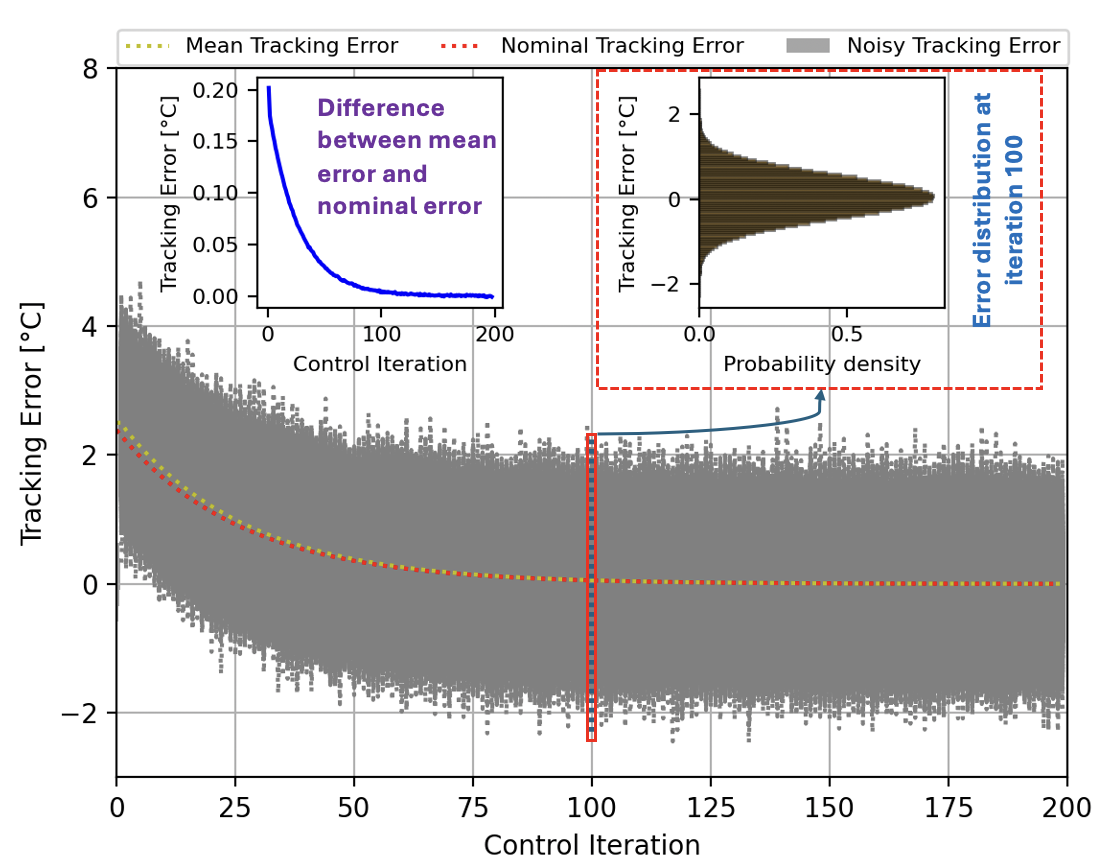}
        \caption{Tracking error at each control iteration in the presence of Gaussian temperature measurement uncertainty}
    \end{subfigure}
    \begin{subfigure}{0.9\linewidth}
        \centering
        \hspace*{-0.6cm}
        \includegraphics[width=1.1\linewidth, trim={0cm 0cm 0cm 0cm}, clip]{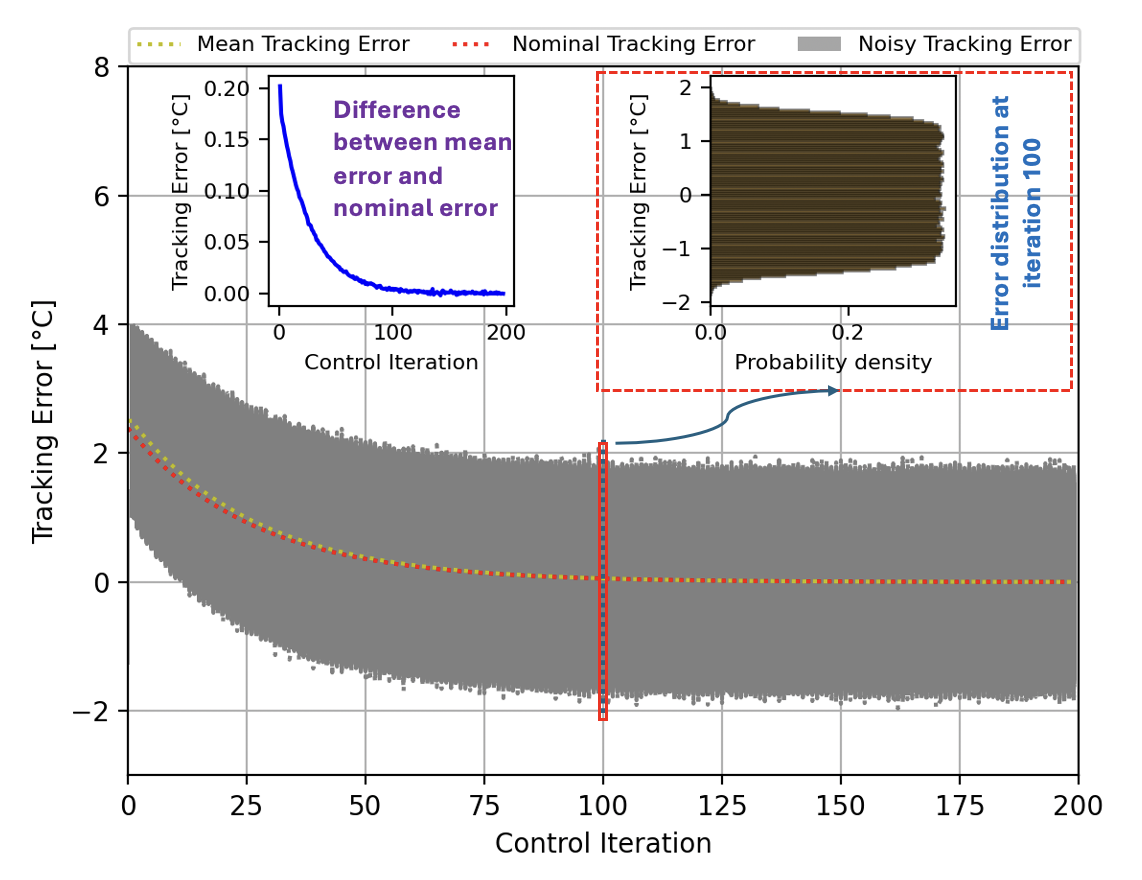}
        \caption{Tracking error at each control iteration in the presence of uniform temperature measurement uncertainty}
    \end{subfigure}
    \caption{Tracking error evolution over control iterations, comparing scenarios with and without temperature measurement uncertainty. Without uncertainty, the nominal error decreases steadily and reaches almost zero after about 150 iterations. In the presence of uncertainty of type of either Gaussian temperature measurement uncertainty $\sim N(0, 0.5)$ (a) and uniform temperature measurement uncertainty $\sim U[-1.5,1.5]$, a visible range of variation appears at each iteration, shown from the first to the last iteration. Both Gaussian and uniform distribution are captured for the tracking error at iteration 100.}~\label{fig:error_history}
\end{figure}

For such a distribution of measured temperature in each control iteration, we can evaluate the distribution of tracking error. We conduct the Monte Carlo simulation for the closed-loop system, which incorporates the heat transfer model described in Equation (4), the heat source outlined in Equation (5), and the PID controller depicted in Equation (6). Figure~\ref{fig:error_history} illustrates the trajectory of tracking errors over different control iterations, considering the presence and absence of uncertainty in the measured temperatures. Figure~\ref*{fig:error_history} shows that without uncertainty, the nominal tracking error decreases steadily and approaches zero after about 150 iterations. However, in the presence of measurement uncertainty, there is a range of variations in the tracking error at each control iteration. To validate the distributional description of the tracking error, we plot its distribution at iteration 100, which shows a Gaussian distribution (Fig.~\ref*{fig:error_history} (a)) and uniform distribution (Fig.~\ref*{fig:error_history} (b)). 

In both Gaussian and uniform measurement uncertainty scenarios, the disparity between the mean of distributions at each control iteration and the nominal error steadily diminishes from approximately 0.2\,\textdegree C to near zero. This trend indicates that as time progresses, the system's behaviour aligns more closely with the expected nominal performance.

\subsection{Uncertainty quantification of tracking error and laser power}
\indent\vspace{-10pt}

To assess the impact of temperature measurement uncertainties on tracking error and laser power, we use mean, standard deviation, kurtosis, skewness, mode and confidence intervals. 
\begin{figure*}[!b]
    \centering
    \captionsetup[subfigure]{oneside,margin={0cm,0.5cm}}
    \begin{subfigure}[t]{0.45\textwidth}
        \centering
        \hspace*{-1cm}
        \includegraphics[width=1.1\linewidth]{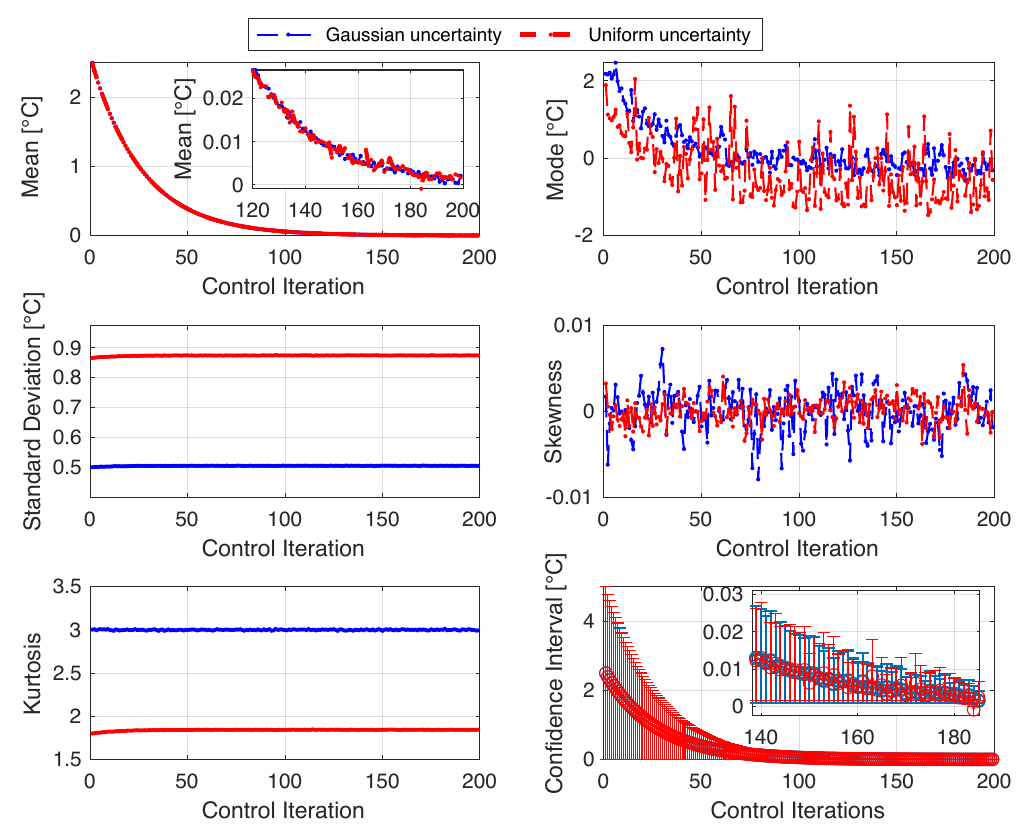}
        \caption{Statistical measures for tracking error distributions over control iterations in the presence of Gaussian measurement uncertainty.}
    \end{subfigure}
    \captionsetup[subfigure]{oneside,margin={1cm,-0.7cm}}
    \begin{subfigure}[t]{0.45\textwidth}
        \centering
        \hspace*{0.2cm}
        \includegraphics[width=1.1\linewidth]{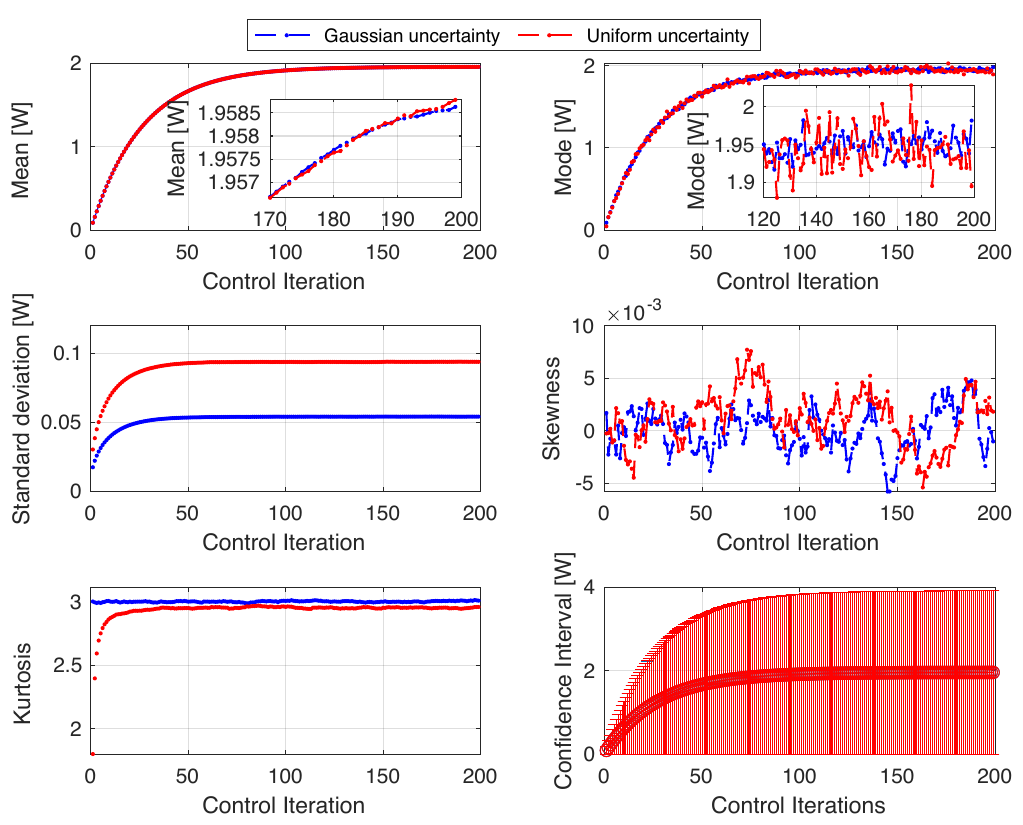}
        % \hspace{1cm} % Adjust the value as needed to move the caption to the right
        \caption{Statistical measures for laser power distributions over control iterations in the presence of Gaussian measurement uncertainty.}
    \end{subfigure}
    \caption{The results depict changes in key metrics during control iterations. The mean converges to a value of 0.002\,\textdegree, indicating consistent behaviour at final control iterations. The standard deviation trend depicts that the tracking error in the presence of uniform measurement uncertainties has larger variability around the mean compared to results in the presence of Gaussian measurement uncertainty. Kurtosis and skewness suggest changes in distribution shape and asymmetry. The mode of error indicates a shift towards lower temperature values. The confidence interval becomes progressively narrower, reflecting increased precision in error distribution estimates over control iterations. Laser power distributions show a systematic increase in mean values over control iterations. Variability, as indicated by standard deviation, fluctuates and stabilises. The distribution of the laser power, whether influenced by uniform or Gaussian measurement uncertainty, tends to be approximately symmetric around its mean. The mode gradually increases and stabilises around 2\,W. The confidence interval widens, reflecting the increasing uncertainty in the measurement estimates as the control iterations progress.}
    \label{fig:error-measures}
\end{figure*}

We use the measures of mean and kurtosis to assess the symmetry of the distribution and to check the expected error values at each control iteration. The standard deviation provides a measure of the dispersion within the data set. Skewness helps to identify outliers. We calculate the mode, the most common value in the data set, and confidence intervals to identify a range of values that are likely to encompass a population parameter with a specified level of confidence.

Figure~\ref*{fig:error-measures} (a) shows statistical measures derived from the tracking error distributions across control iterations in the presence of Gaussian and uniform temperature measurement uncertainty. The mean values start at about 2.5\,\textdegree C and gradually decrease to about 0.002\,\textdegree C in the final iterations. The standard deviation with uniform temperature measurement uncertainty varies around 0.9\,\textdegree C, while for Gaussian temperature measurement uncertainty it varies around 0.5\,\textdegree C. It indicates that the tracking error in the presence of uniform temperature measurement uncertainty has a greater variability around the mean compared to the results in the presence of Gaussian temperature measurement uncertainty.

In the presence of uniform temperature measurement uncertainty, kurtosis remains stable, ranging from 2.99\,\textdegree C to 3.01\,\textdegree C, indicating a consistent distribution shape. However, with Gaussian temperature measurement uncertainty, kurtosis varies between 1.83\,\textdegree C and 1.85\,\textdegree C, showing a more peaked distribution with heavier tails. 

The mode starts at approximately 2.5\,\textdegree C and declines over time, suggesting a shift toward lower temperatures. For uniform temperature measurement uncertainty, mode variations range from -1.2\,\textdegree C to 1.6\,\textdegree C, while with Gaussian temperature measurement uncertainty, they're narrower, from -0.3\,\textdegree C to 0.4\,\textdegree C.

The skewness is between -0.007\,\textdegree C and 0.007\,\textdegree C and -0.004\,\textdegree C and 0.004\,\textdegree C for all iterations in the presence of Gaussian and uniform temperature measurement uncertainties, respectively. The results indicate that the error distribution is approximately symmetric with no significant skewness in either direction. The slight differences in the range of skewness between the two types of measurement uncertainties suggest that Gaussian temperature measurement uncertainty may produce slightly wider distributions than uniform temperature measurement uncertainty. 

The confidence interval in the presence of both uniform and Gaussian temperature measurement uncertainties, calculated at the $95\%$ confidence level, begins between zero and 5\,\textdegree C in the early iterations and gradually decreases to a narrower interval of about zero to 0.001\,\textdegree C, indicating a reduction in the variability of the tracking error over the iterations.

In Figure~\ref*{fig:error-measures} (b), the mean of the laser power distributions with either uniform or Gaussian temperature measurement uncertainties shows a gradual increase over the control iterations, progressing from zero to about 2\,W. This suggests a systematic trend towards higher values in the laser power distribution as the control process evolves. The standard deviation initially increases from zero to 0.054\,W and 0.094\,W for Gaussian and uniform temperature measurement uncertainties, respectively. After 70 iterations, there is no significant variation. This indicates that the dispersion of the laser power values initially increases with the number of iterations. This increase then stabilises after a certain number of iterations, indicating a convergence of the variability of the laser power values over time, regardless of the type of temperature measurement uncertainty.

The skewness and kurtosis for both uniform and Gaussian temperature measurement uncertainties vary around zero and 3 respectively. The distribution of laser power, whether influenced by uniform or Gaussian temperature measurement uncertainties, tends to be approximately symmetric around its mean (skewness around zero) and has tails similar to those of a normal distribution (kurtosis around 3). This implies that the temperature measurement uncertainty does not significantly distort the shape of the distribution in terms of its symmetry or tail behaviour. 

The mode in the presence of both uniform and Gaussian temperature measurement uncertainties shows a gradual increase, reaching a stable value of about 2\,W after about 70 iterations. This indicates a shift in the most frequently occurring value, eventually stabilising around 2\,W in the later stages of the control iterations. The larger modes for uniform temperature measurement uncertainty mean that laser powers are more concentrated around the mean values compared to the Gaussian temperature measurement uncertainty. The confidence interval starts at zero and gradually increases, approaching intervals between zero and 4\,W in the final iterations. This means there is increasing variability in the laser power over control iterations.
    
\subsection{Comparison between Monte Carlo simulation and Laplace}
\indent\vspace{2pt}

In this section, we demonstrate that we can accurately estimate the distribution of tracking error and laser power using Laplace. This method allows us to assess uncertainty just as precisely as the Monte Carlo simulation but with less computational effort. With Laplace, we can determine the distribution of error and laser power at each control step without having to simulate the entire process multiple times, as Monte Carlo does, to see how sensitive the system is to temperature measurement uncertainty.

We compare Laplace and Monte Carlo simulation to understand the balance between computational efficiency and accuracy in estimating tracking error distributions. To measure how different the results from each method are compared to the true distribution (obtained from a Monte Carlo simulation with one million iterations), we use Wasserstein distance metric. We repeat the entire analysis 30 times for both Laplace and Monte Carlo under both uniform or Gaussian temperature measurement uncertainties to calculate the average and standard deviation of the Wasserstein distance and runtime.

\begin{figure*}[!t]
    \centering
    \begin{subfigure}[t]{0.45\textwidth}
        \centering
        \hspace*{-0.3cm}
        \includegraphics[width=1.1\linewidth]{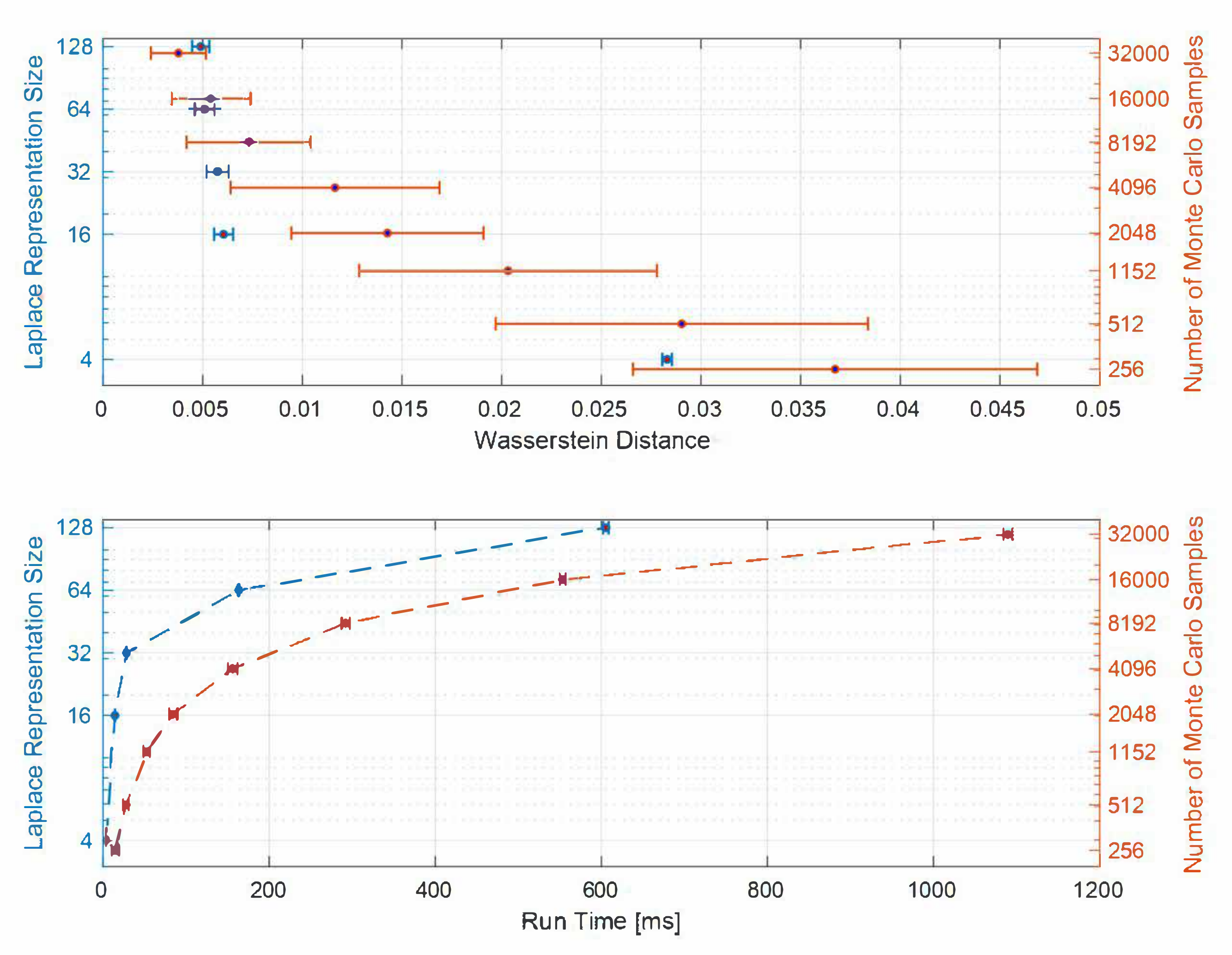}        
        \caption{Wasserstein distance between the Monte Carlo simulation and the Ground truth distribution, and between Laplace and the Ground truth distribution (top figure). Run time needed to generate error distribution for Monte Carlo simulation and Laplace under the influence of Gaussian temperature measurement uncertainty (bottom figure).}
    \end{subfigure}
    \hfill 
    \begin{subfigure}[t]{0.45\textwidth}
        \centering
        \hspace*{-0.75cm}
        \includegraphics[width=1.14\linewidth]{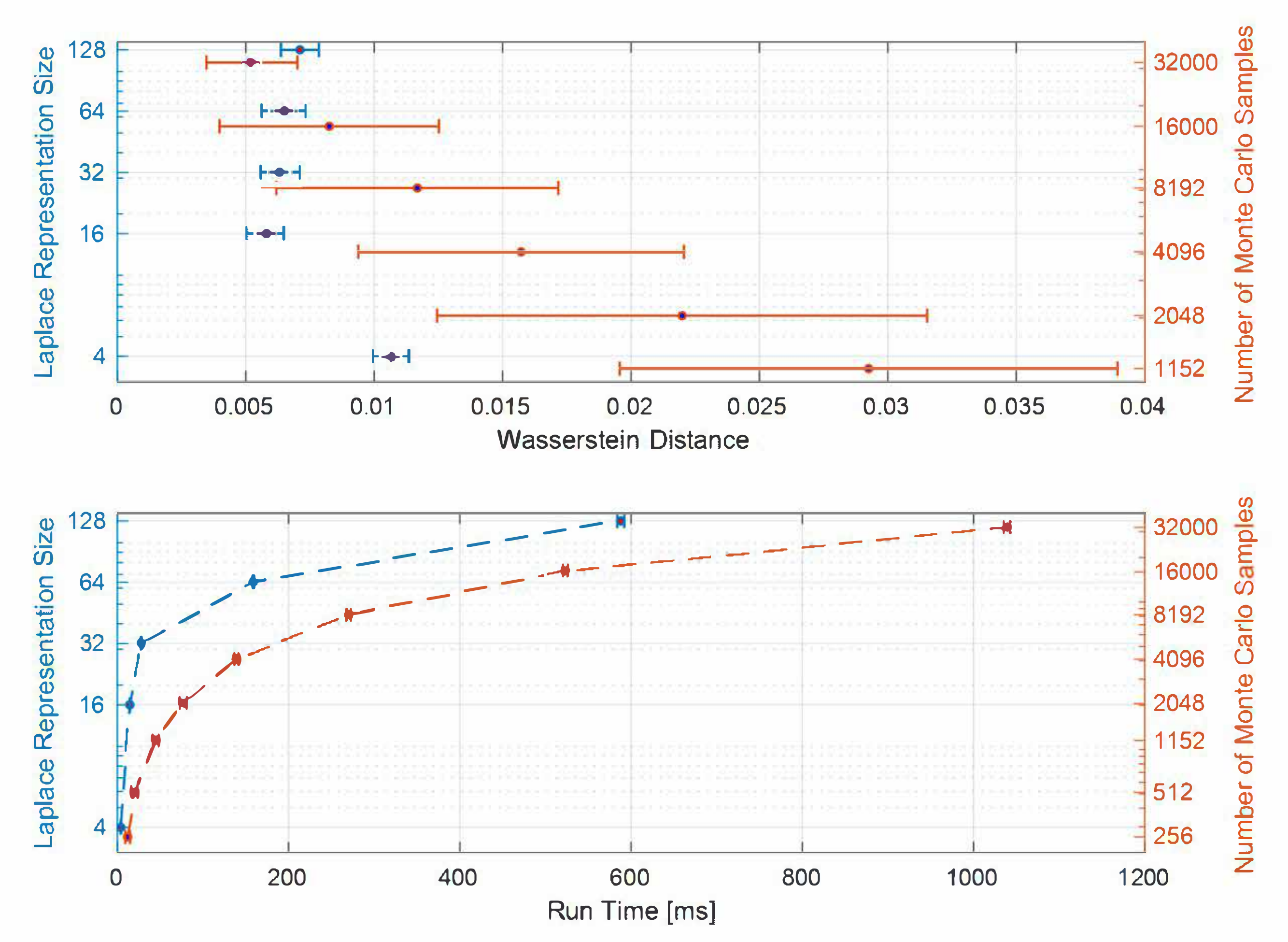}        
        \caption{Wasserstein distance between the Monte Carlo simulation and the Ground truth distribution, and between Laplace and the Ground truth distribution (top figure). Run time needed to generate error distribution for Monte Carlo simulation and Laplace under the influence of uniform measurement temperature measurement uncertainties (bottom figure).}
    \end{subfigure}
    \caption{In the presence of Gaussian temperature measurement uncertainty, the accuracy of error distribution estimation using Laplace with representation size of 32 is higher than Monte Carlo with 16000 iterations. Under uniform temperature measurement uncertainty, Laplace with representation size of 16 can achieve better accuracy than Monte Carlo with 32000 iterations. Looking at the runtime results, Laplace is 19$\times$ and 71$\times$ faster than Monte Carlo for the same accuracy under uniform and Gaussian temperature measurement uncertainties, respectively.}\label{fig:Wasserstein}
\end{figure*}

\begin{table*}[!b]
    \centering
    \caption{Comparison of Wasserstein distance and run time for steady-state tracking error distributions between Laplace with different representation sizes and traditional Monte Carlo simulation with different numbers of samples. For both algorithms, we repeat the whole process 30 times to obtain the mean and standard deviation of the Wasserstein distance and run time parameters. To achieve better accuracy than Laplace under Gaussian and uniform temperature measurement uncertainty, the traditional Monte Carlo simulation requires at least 16000 and 32000 number of iterations, respectively. This means that for the same level of accuracy, Laplace is approximately 19$\times$ and 71$\times$ faster than Monte Carlo under Gaussian and uniform temperature measurement uncertainties, respectively.}
    \begin{tabular}{cccccc}        
        \toprule
        \textbf{Core} & \textbf{Representation Size/Number of Samples} & \textbf{Wasserstein Distance} & \textbf{Measurement Uncertainty Type} & \textbf{Run time (ms)} \\
        \midrule
        Laplace & 4 & $0.02830 \pm 0.00024$ & Gaussian & $3.974 \pm 0.062$  \\
        Laplace & 16 & $0.00605 \pm 0.00048$ & Gaussian & $14.342 \pm 0.081$  \\
        \textbf{Laplace} & \textbf{32} & $\boldsymbol{0.00580 \pm 0.00055}$  & \textbf{Gaussian} & $\boldsymbol{29.509 \pm 0.135}$  \\
        Laplace & 64 & $0.00514 \pm 0.00050$ & Gaussian & $164.398 \pm 0.741$ \\
        Laplace & 128 & $0.00490 \pm 0.00043$ & Gaussian & $605.530 \pm 3.266$  \\
        \bottomrule
        \addlinespace[0.5ex]
        Traditional Monte Carlo & 256 & $0.03673 \pm 0.01014$ & Gaussian & $15.358 \pm 4.687$  \\
        Traditional Monte Carlo & 512 & $0.02904 \pm 0.00934$ & Gaussian & $28.859 \pm 3.490$  \\
        Traditional Monte Carlo & 1152 & $0.02032 \pm 0.00747$ & Gaussian & $53.645 \pm 4.023$  \\
        Traditional Monte Carlo & 2048 & $0.01427 \pm 0.00482$ & Gaussian & $85.686 \pm 4.886$   \\
        Traditional Monte Carlo & 4096 & $0.01164 \pm 0.00524$ & Gaussian & $157.080 \pm 5.790$  \\
        Traditional Monte Carlo & 8192 & $0.00732 \pm 0.00313$ & Gaussian & $293.571 \pm 4.891$ \\
        \textbf{Traditional Monte Carlo} & \textbf{16000} & $\boldsymbol{0.00544 \pm 0.00200}$ & \textbf{Gaussian} & $\boldsymbol{554.747 \pm 3.765}$  \\
        Traditional Monte Carlo & 32000 & $0.00378 \pm 0.00138$ & Gaussian & $1089.929 \pm 5.339$ \\
        \midrule
        Laplace & 4 &  $0.01069 \pm 0.00070$  & Uniform & $3.946 \pm 0.067$ \\
        \textbf{Laplace} & \textbf{16} & $\boldsymbol{0.00578 \pm 0.00071}$ & \textbf{Uniform} & $\boldsymbol{14.604 \pm 0.092}$  \\
        Laplace & 32 & $0.00637 \pm 0.00076$ & Uniform & $28.722 \pm 0.117$  \\
        Laplace & 64 & $0.00651 \pm 0.00083$ & Uniform & $159.782 \pm 0.748$ \\
        Laplace & 128 & $0.00711 \pm 0.00074$ & Uniform & $587.920 \pm 4.121$  \\
        \bottomrule
        \addlinespace[0.5ex]
        Traditional Monte Carlo & 256 & $0.06340 \pm 0.02288$ & Uniform & $12.079 \pm 2.812$  \\
        Traditional Monte Carlo & 512 & $0.04056 \pm 0.01762$ & Uniform & $21.329 \pm 3.884$  \\
        Traditional Monte Carlo & 1152 & $0.02925 \pm 0.00969$ & Uniform & $45.912 \pm 3.291$  \\
        Traditional Monte Carlo & 2048 & $0.02199 \pm 0.00954$ & Uniform &  $76.898 \pm 4.226$  \\
        Traditional Monte Carlo & 4096 & $0.01572 \pm 0.00634$ & Uniform & $140.268 \pm 2.313$  \\
        Traditional Monte Carlo & 8192 & $0.01168 \pm 0.00549$ & Uniform & $271.253 \pm 2.542$ \\
        Traditional Monte Carlo & 16000 & $0.00825 \pm 0.00427$ & Uniform & $523.482 \pm 2.837$ \\
        \textbf{Traditional Monte Carlo} & \textbf{32000} & $\boldsymbol{0.00524 \pm 0.00177}$ & \textbf{Uniform} & $\boldsymbol{1040.454 \pm 3.950}$ \\
        \bottomrule        
    \end{tabular}
    \label{table:mc-laplace}
\end{table*}

\begin{figure*}[!b]
    \centering
    \begin{subfigure}{0.2\textwidth}
    \hspace*{-0.2cm}
    \includegraphics[width=1.1\linewidth]{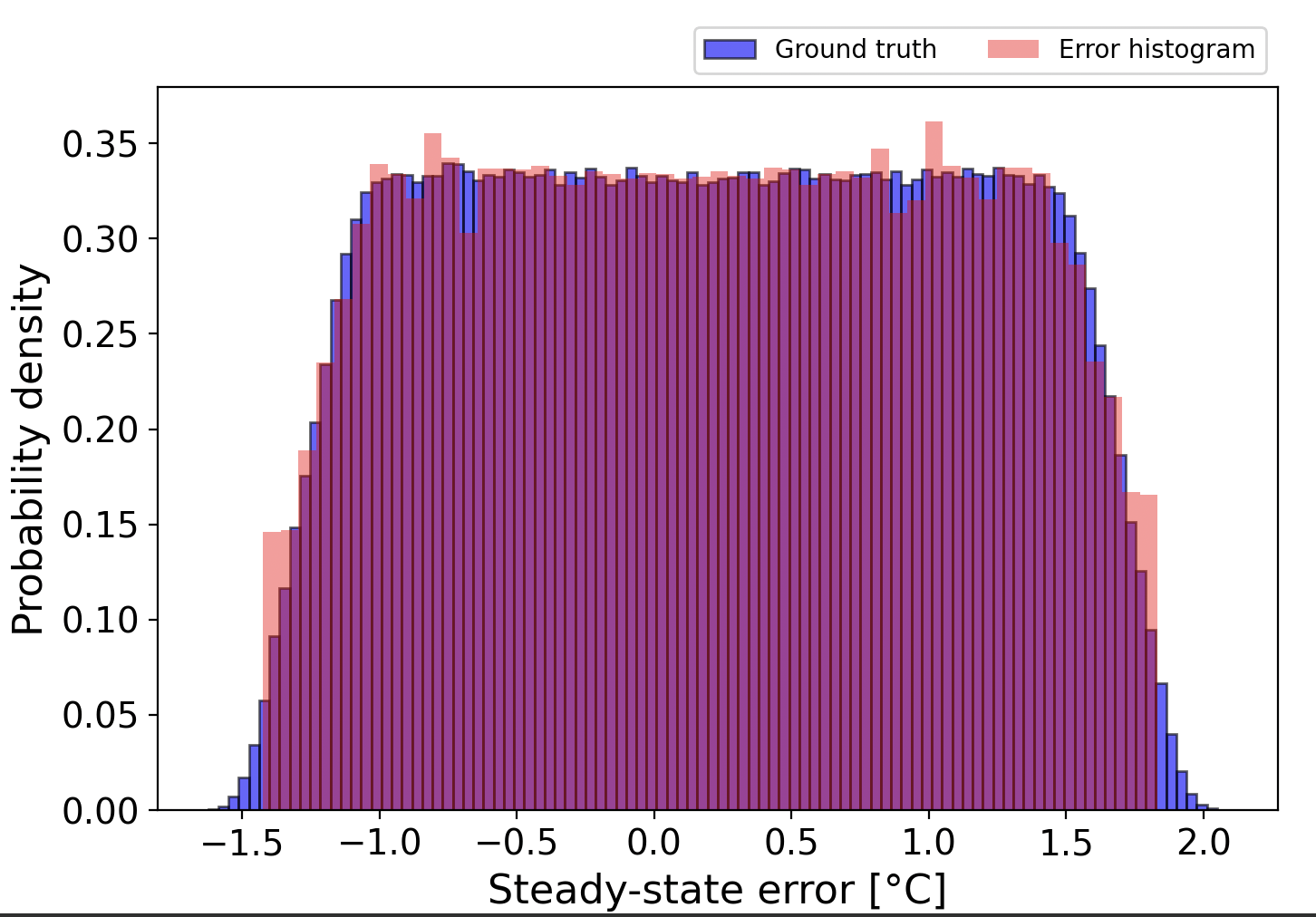}
    \caption{Laplace with Representation size 16, Wasserstein distance mean: 0.00578.}
    \end{subfigure}%
    \hfill
    \begin{subfigure}{0.2\textwidth}
    \includegraphics[width=1.1\linewidth]{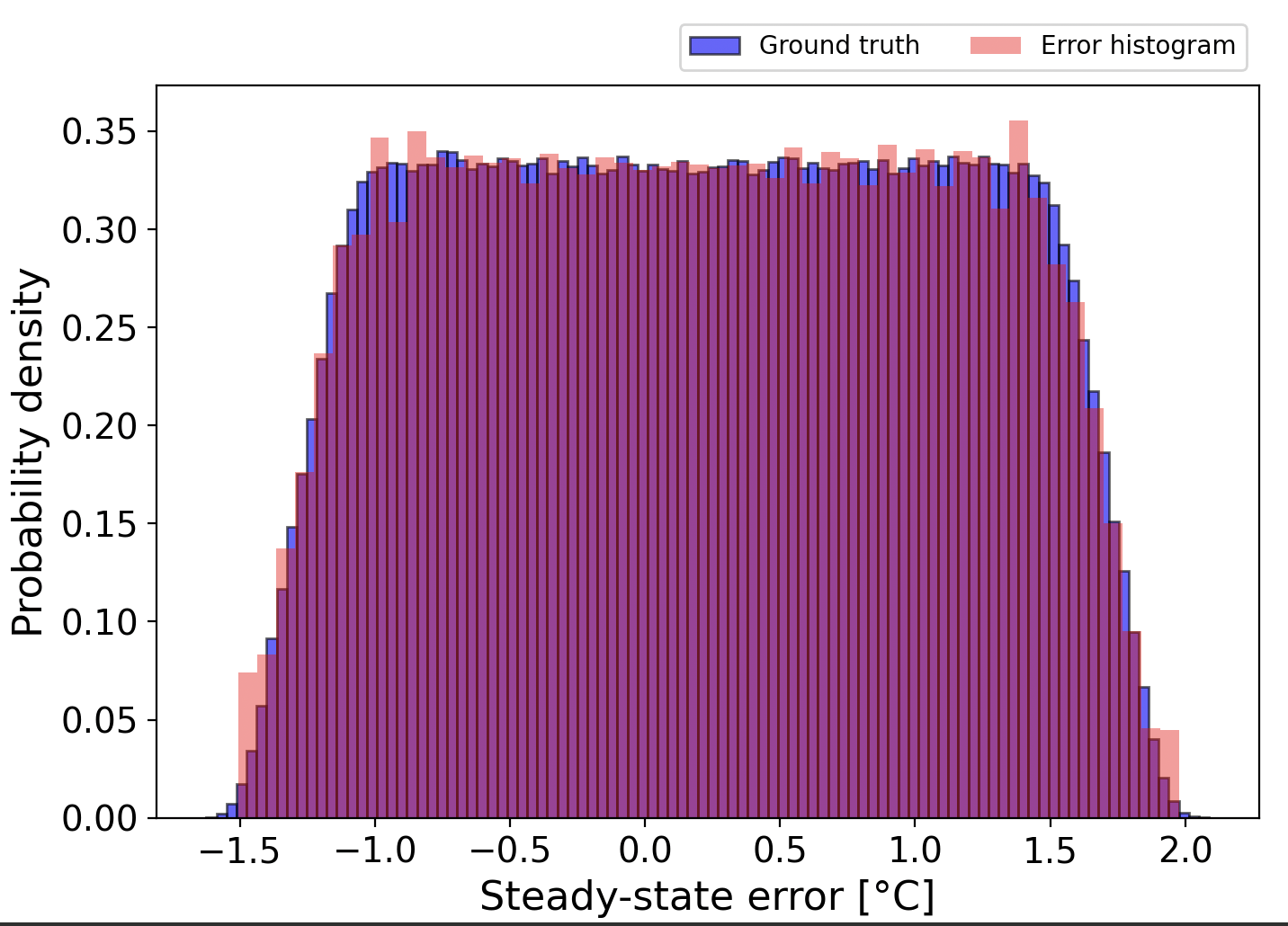}
    \caption{Laplace with Representation size: 32, Wasserstein distance mean: 0.00637.}
    \end{subfigure}%
    \hfill
    \begin{subfigure}{0.2\textwidth}
    \includegraphics[width=1.1\linewidth]{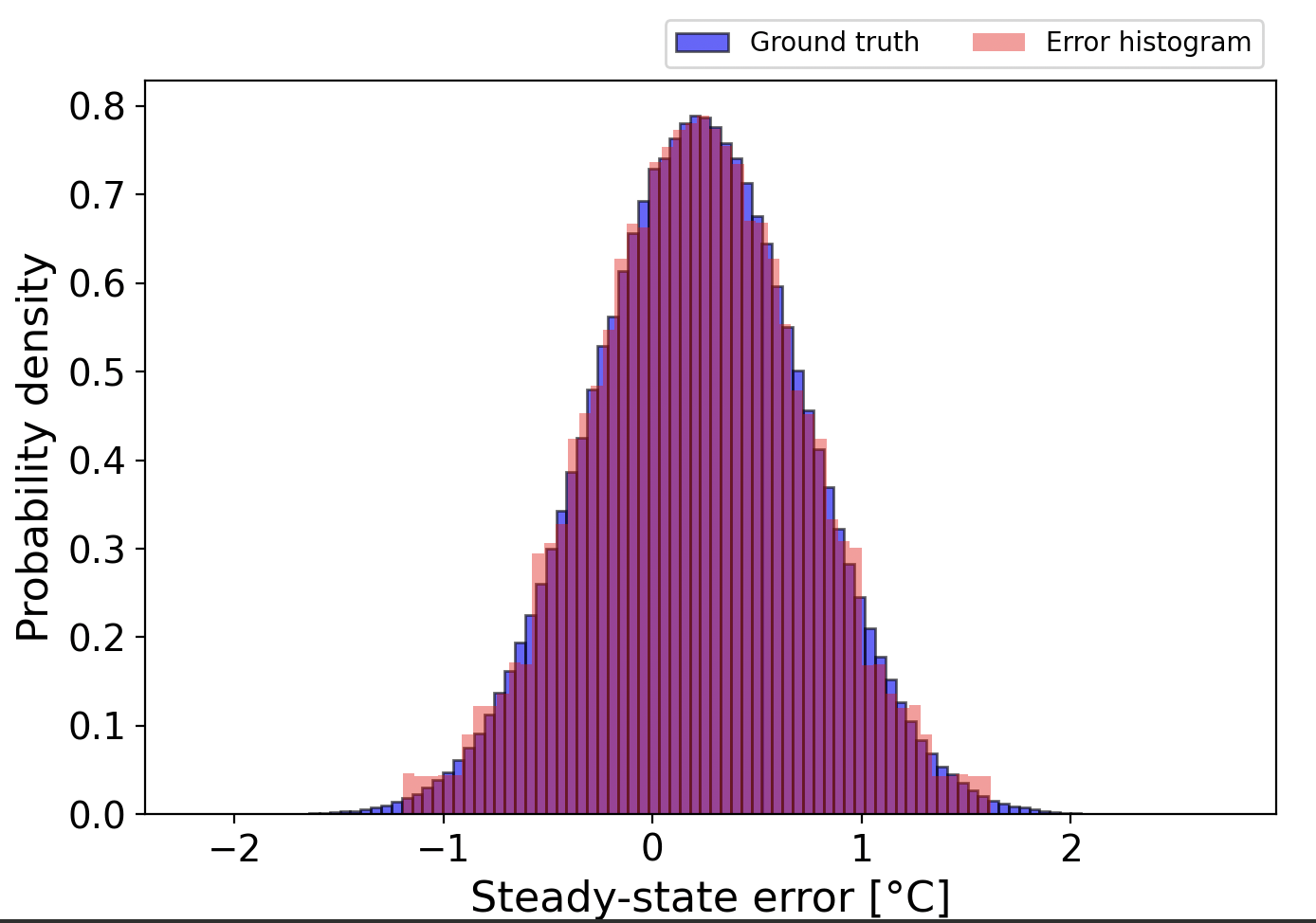}
    \caption{Laplace with Representation size: 16, Wasserstein distance mean: 0.00605.}
    \end{subfigure}%
    \hfill
    \begin{subfigure}{0.2\textwidth}
    \includegraphics[width=1.1\linewidth]{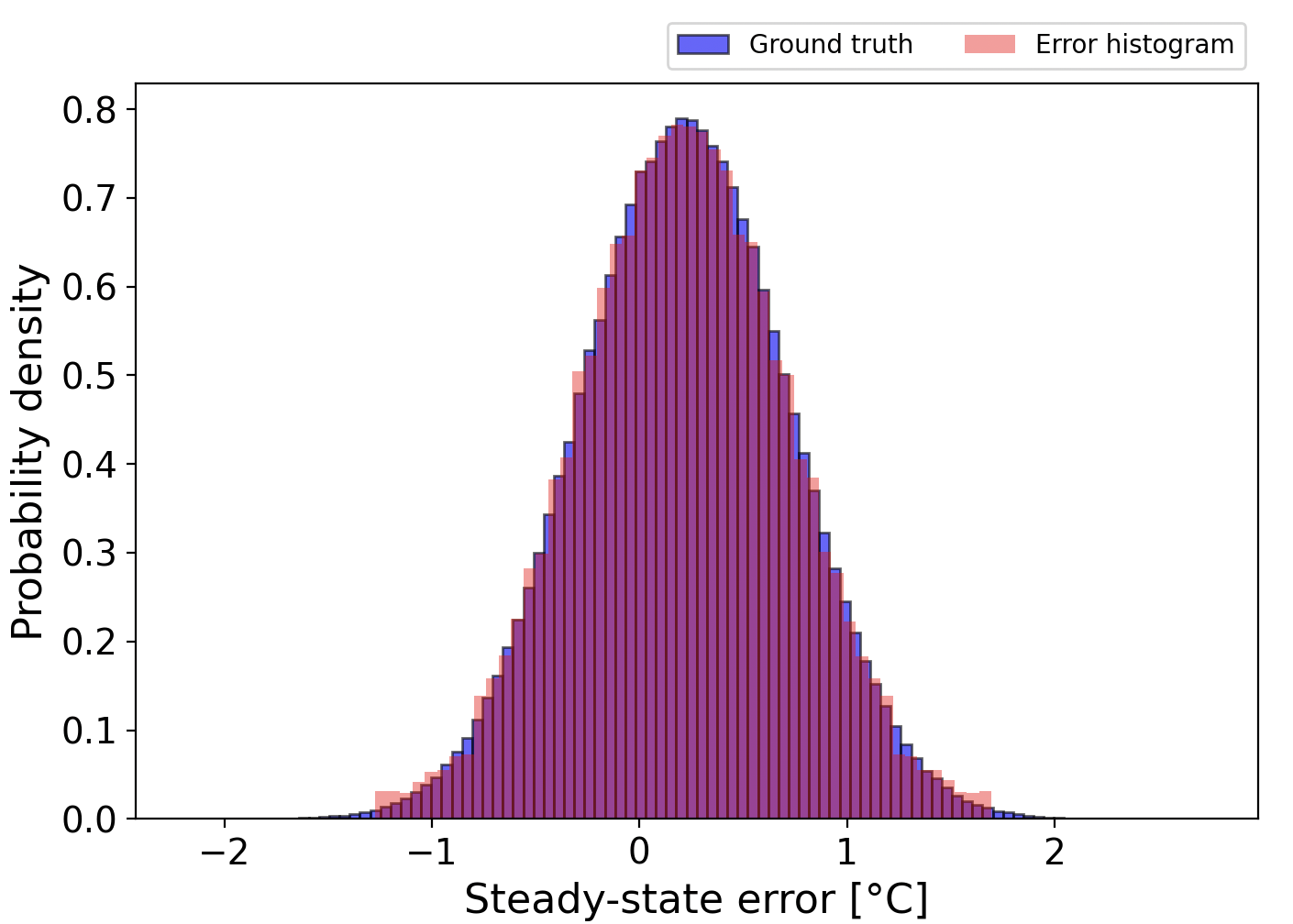}
    \caption{Laplace with Representation size: 32, Wasserstein distance mean: 0.00580.}
    \end{subfigure}
    \begin{subfigure}{0.2\textwidth}
    \includegraphics[width=1.1\linewidth]{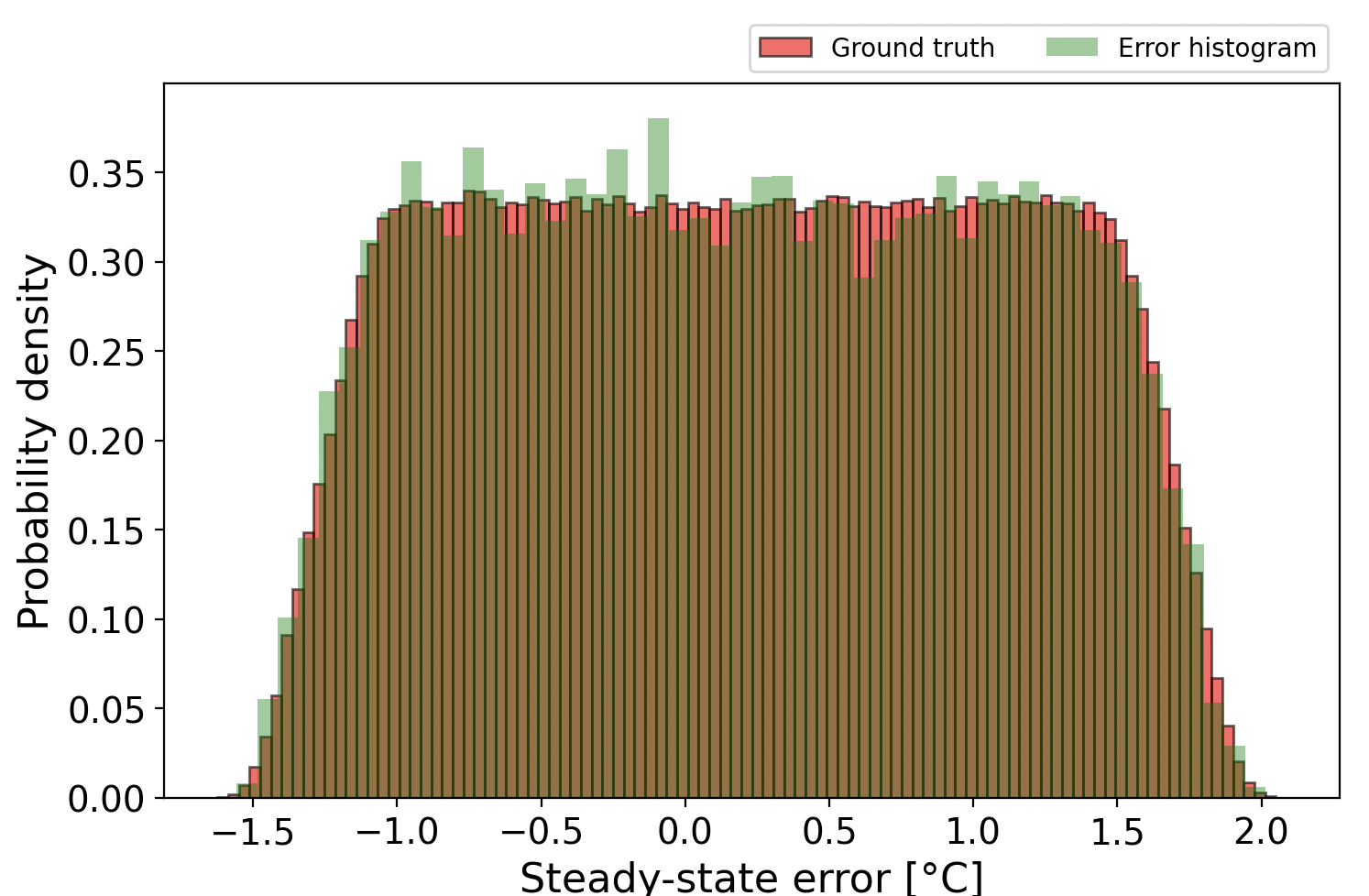}
    \caption{Monte Carlo simulation. Number of Monte Carlo iterations: 16000, Wasserstein distance mean: 0.00825.}
    \end{subfigure}%
    \hfill
    \begin{subfigure}{0.2\textwidth}
    \includegraphics[width=1.1\linewidth]{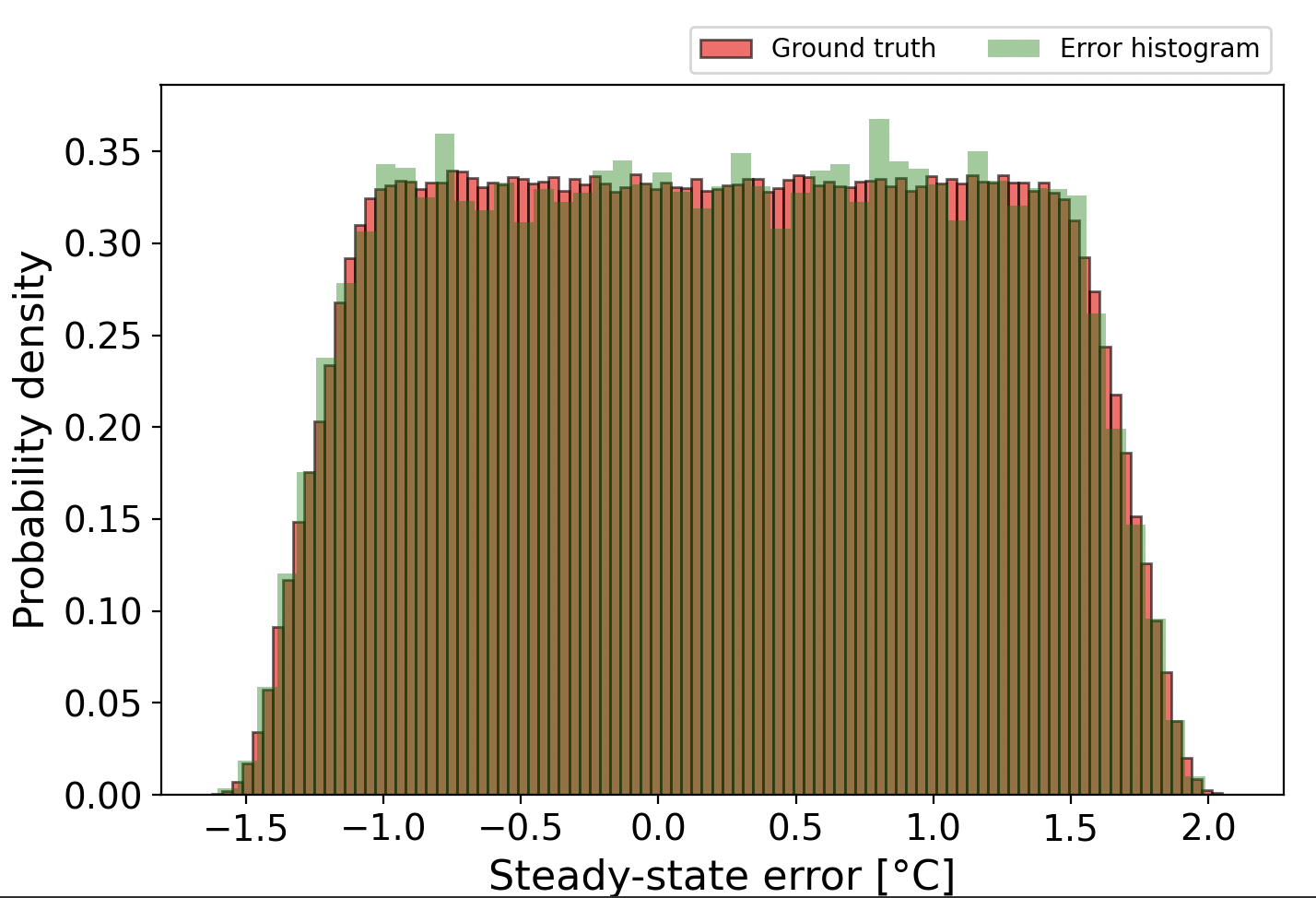}
    \caption{Monte Carlo simulation. Number of Monte Carlo iterations: 32000, Wasserstein distance mean: 0.00524.}
    \end{subfigure}%
    \hfill
    \begin{subfigure}{0.2\textwidth}
    \includegraphics[width=1.1\linewidth]{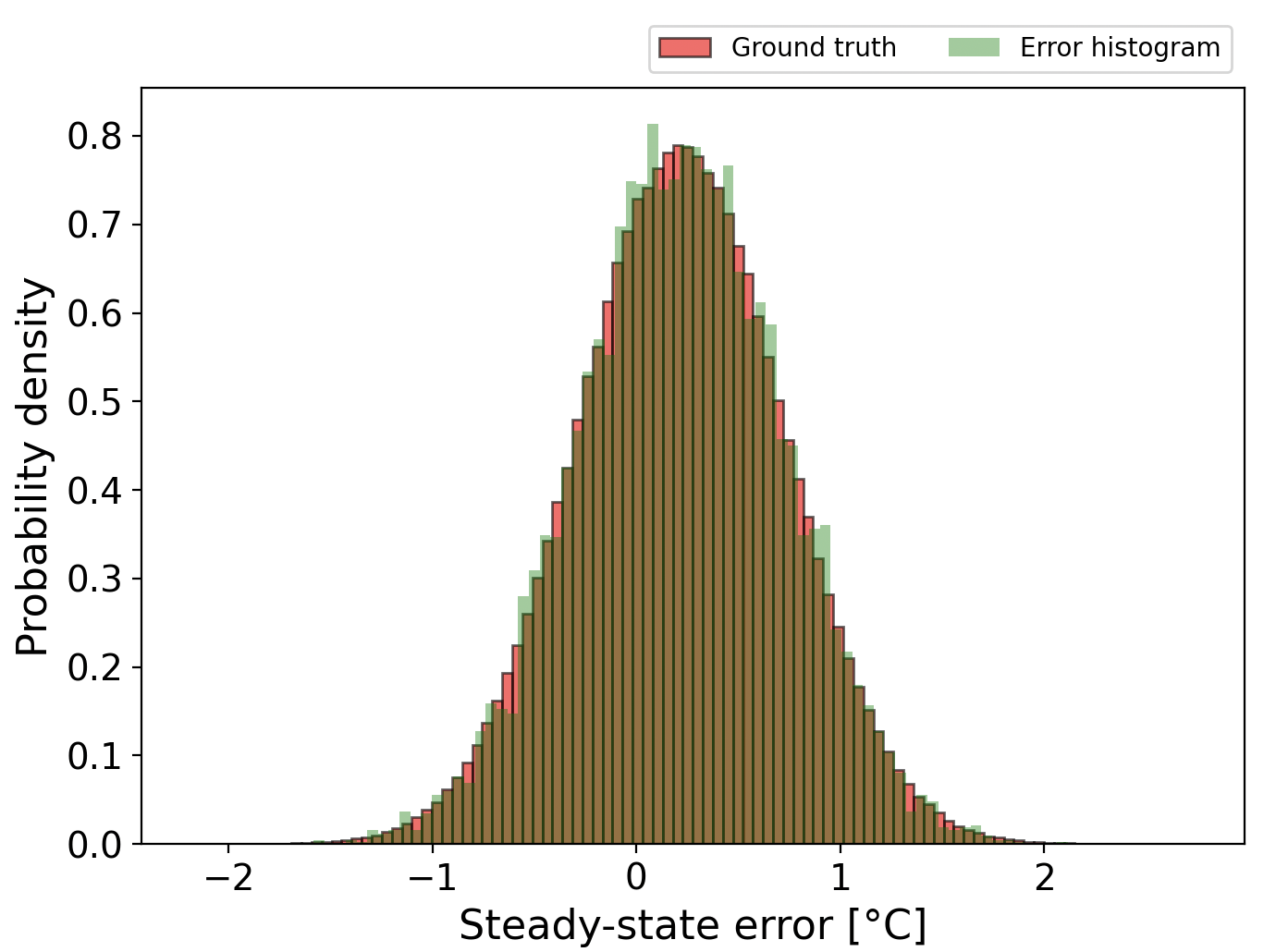}
    \caption{Monte Carlo simulation. Number of Monte Carlo iterations: 8192, Wasserstein distance mean: 0.00732.}
    \end{subfigure}%
    \hfill
    \begin{subfigure}{0.2\textwidth}
    \includegraphics[width=1.1\linewidth]{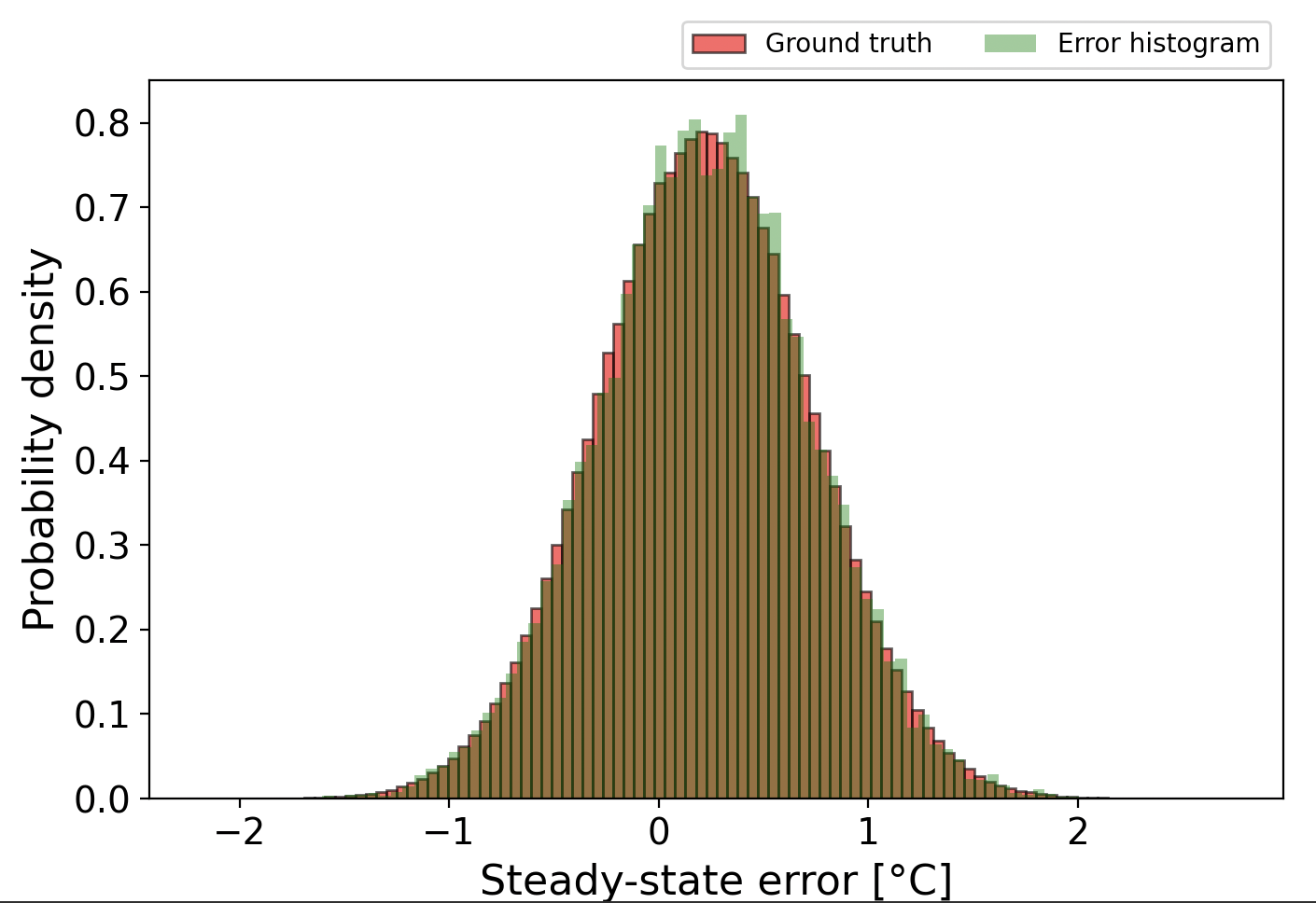}
    \caption{Monte Carlo simulation. Number of Monte Carlo iterations: 16000, Wasserstein distance mean: 0.00544.}
    \end{subfigure}
    \caption{Comparative analysis of convergence challenges in Monte Carlo simulation and Laplace for estimating steady-state tracking error distribution. The figures highlight the performance variation of each method in approximating the steady-state error distribution.}\label{fig: mc-laplace-hist}
\end{figure*}

Table~\ref{table:mc-laplace} shows that under Gaussian temperature measurement uncertainty, Laplace performs consistently well across different representation sizes (except for size 4), with Wasserstein distance averages ranging from 0.00605 to 0.00490. This indicates that Laplace provides stable and reliable estimates of the error distribution. 

The traditional Monte Carlo simulation shows a noticeable improvement in accuracy as the number of samples increases, with the Wasserstein distance decreasing significantly from 0.03673 with 256 number of Monte carlo iterations to 0.00378 with 32,000 number of Monte Carlo iterations.

This indicates the effectiveness of Monte Carlo simulation in capturing the true distribution with greater precision, especially at larger number of Monte Carlo iterations. Laplace with a representation size of 32 can achieve a same level of accuracy compared to the traditional Monte Carlo simulation with 16,000  number of iterations. Examining the required runtime for each method reveals that Laplace is approximately 17 times faster than Monte Carlo simulation to achieve the same level of accuracy with Gaussian temperature measurement uncertainty.

The results in Figure~\ref*{fig:Wasserstein} depict a comparison between Laplace and traditional Monte Carlo simulation methods for estimating steady-state tracking error distributions under both Gaussian and uniform temperature measurement uncertainties. In Figure~\ref*{fig:Wasserstein} (a), representing Gaussian temperature measurement uncertainty, the top subplot demonstrates their performance based on Wasserstein distance, while the bottom subplot displays their execution run time. Similarly, Figure~\ref*{fig:Wasserstein} (b) presents the comparison under uniform temperature measurement uncertainty.

Under uniform temperature measurement uncertainty, Laplace maintains relatively stable Wasserstein distances ranging from 0.00578 with a representation size of 16 to 0.00711 with a size of 128. In contrast, traditional Monte Carlo simulation achieves low Wasserstein distances of 0.00524 with a relatively large number of Monte Carlo iterations. To exceed Laplace in accuracy, Monte Carlo simulation requires 32000 iterations, resulting in a minimum runtime of about 1040 milliseconds. 

At the same level of accuracy, Laplace estimates the steady-state error distribution with a representation size of 16 in about 14.6 milliseconds. This implies that Laplace is approximately 71 times faster than Monte Carlo under uniform temperature measurement uncertainty. It's worth noting that the Laplace approximation requires running the entire control process only once, whereas Monte Carlo simulation necessitates simulating the closed-loop system multiple times.

Examining the plots in Figure~\ref*{fig:Wasserstein}, it's evident that both Laplace and Monte Carlo simulation exhibit different behaviours under varying temperature measurement uncertainty distributions. Increasing the representation size in Laplace leads to smaller Wasserstein distances but longer run times. Laplace requires approximately 30 milliseconds and 15 milliseconds to achieve comparable accuracy to Monte Carlo simulation in the presence of Gaussian and uniform temperature measurement uncertainties, respectively. 

Monte Carlo simulation requires a minimum of 16,000 iterations, corresponding to a runtime of 554 milliseconds under Gaussian temperature measurement uncertainty and 32,000 iterations, corresponding to 1,040 milliseconds under uniform temperature measurement uncertainty, to surpass Laplace in accuracy. This highlights that traditional Monte Carlo simulation is significantly slower than Laplace at the same accuracy level. Laplace is about 19 times faster than Monte Carlo under Gaussian temperature measurement uncertainty and 71 times faster under uniform temperature measurement uncertainty. This illustrates that Laplace significantly outperforms Monte Carlo in terms of computational efficiency while achieving comparable accuracy. The results presented in Figure \ref{fig: mc-laplace-hist} offer a comparative analysis of the convergence challenges encountered in Monte Carlo simulation and Laplace when estimating steady-state tracking error distributions. Each subfigure depicts the performance variation of these methods in approximating the steady-state error distribution under different conditions. 

The histograms visualize the distribution of tracking errors obtained from each method, with captions indicating the representation size or number of samples used, along with the corresponding Wasserstein distance. In the case of Laplace, as the representation size increases, there is a trend of decreasing Wasserstein distance, suggesting improved accuracy in approximating the true distribution. Conversely, for Monte Carlo simulation, variations in the number of Monte Carlo iterations result in differing Wasserstein distances, indicating the impact of number of Monte Carlo iterations on accuracy. The results in Figure \ref{fig: mc-laplace-hist} underscore the trade-offs between computational efficiency and accuracy inherent in these two methods. While Laplace offers faster convergence with smaller representation sizes, Monte Carlo simulation achieves higher accuracy but requires significantly larger number of Monte Carlo iterations, leading to longer computation times.

\section{Conclusions}
\label{section:conclusions}
In this study, we examined how measurement uncertainty in temperatures impacts the performance of a model-based laser power control system in an SLS 3D printer. Through experimental setup, we assessed the range of measurement uncertainty and the accuracy of a thermal camera for temperature capture. Employing Monte Carlo simulation, we represented tracking error and laser power as distributions at each control iteration. We analyzed tracking error and laser power behaviour over iterations under uniform and Gaussian temperature measurement uncertainties, using metrics like mean, standard deviation, mode, kurtosis, skewness, and confidence interval. Comparing two distribution estimation methods, Monte Carlo simulation and Laplace, in terms of Wasserstein distance and execution run time, we found Laplace to be 17 times and 71 times faster than Monte Carlo simulation under Gaussian and uniform temperature measurement uncertainties, respectively, while maintaining the same accuracy level. In a closed-loop laser power system, fast operation is crucial for optimal sintering. Laplace, with its fast runtime and accurate tracking error distribution estimation, emerges as an effective solution for laser power control systems.

\acknow{}

\showacknow % Display the acknowledgments section

% \pnasbreak splits and balances the columns before the references.
% If you see unexpected formatting errors, try commenting out this line
% as it can run into problems with floats and footnotes on the final page.
% \pnasbreak

\vspace{0.25in}

% Bibliography
\bibliography{Biblo}

\onecolumn
\appendix
\twocolumn

\section{Appendix} 
\label{section:appendix}

We use Monte Carlo simulation to estimate tracking error and laser power distributions for each control iteration. We produce 1 million samples of tracking error by simulating the control system with 200 control iterations, during which we sample the temperature measurement noise at each time step. At each control iteration, the control system receives the sampled temperature, calculates the tracking error, adjusts the laser power, and applies it to the thermal model. We simulate this process 1 million times, with 200 time steps per simulation. At each step, we sample the easurement noise and use it to update the tracking error. We then compute the mean and standard deviation to determine the sensitivity of the tracking error and laser power to the measurement uncertainty in the temperatures. The analysis considers two types of measurement uncertainty: uniform measurement uncertainty $\varepsilon_u\sim U[-1.5, 1.5]$ and Gaussian measurement uncertainty $\varepsilon_g\sim N(0, 0.5)$.
\begin{figure}[!b]
    \centering
    \begin{subfigure}{0.9\linewidth}
        \centering
        \includegraphics[width=\linewidth, trim={0cm 0cm 0cm 0cm}, clip]{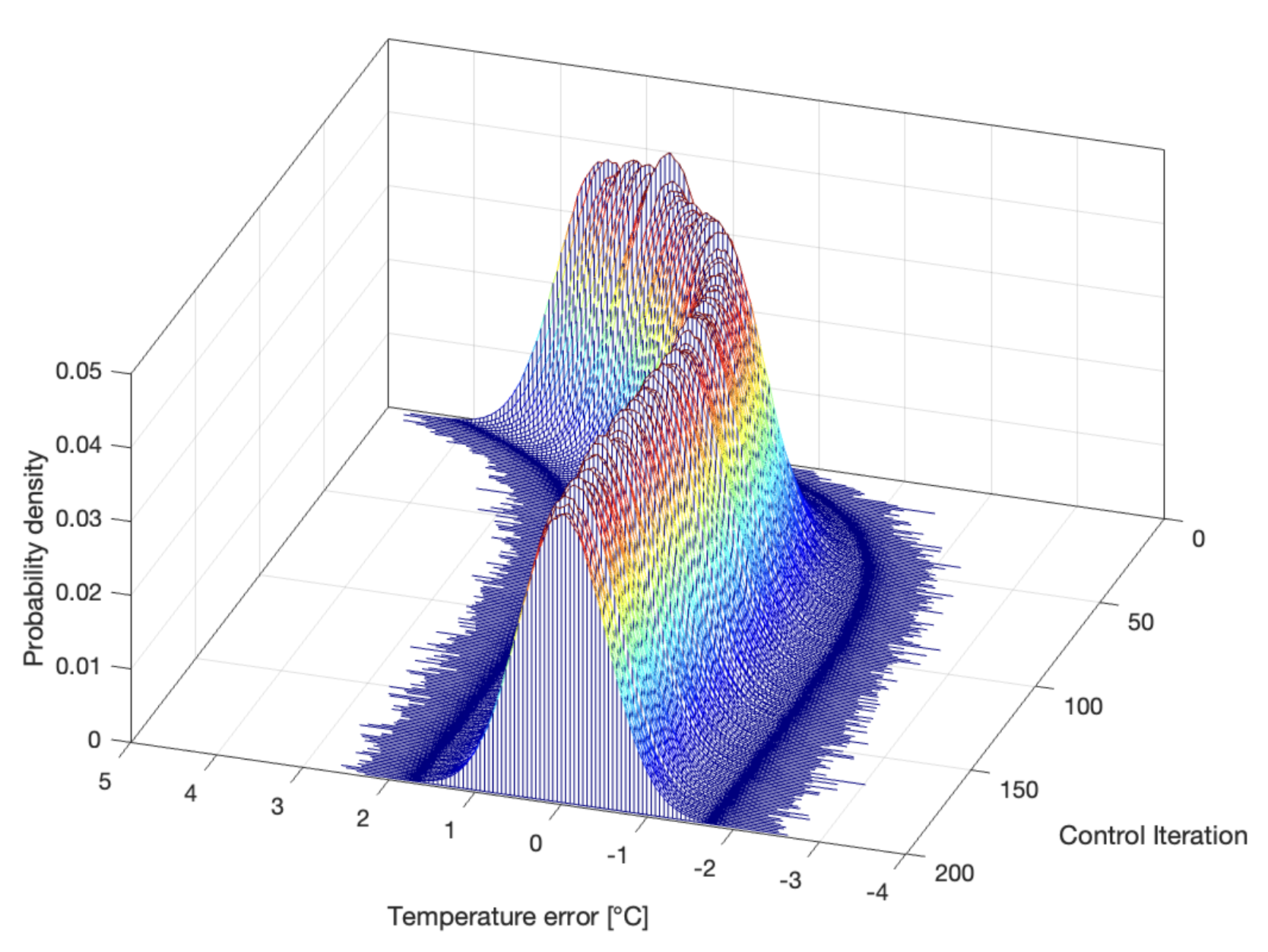}
        \caption{3D view of the tracking error distributions at each control iteration in the presence of Gaussian measurement uncertainty.}
    \end{subfigure}
    \begin{subfigure}{0.95\linewidth}
        \centering
        \includegraphics[width=\linewidth, trim={0cm 0cm 0cm 0cm}, clip]{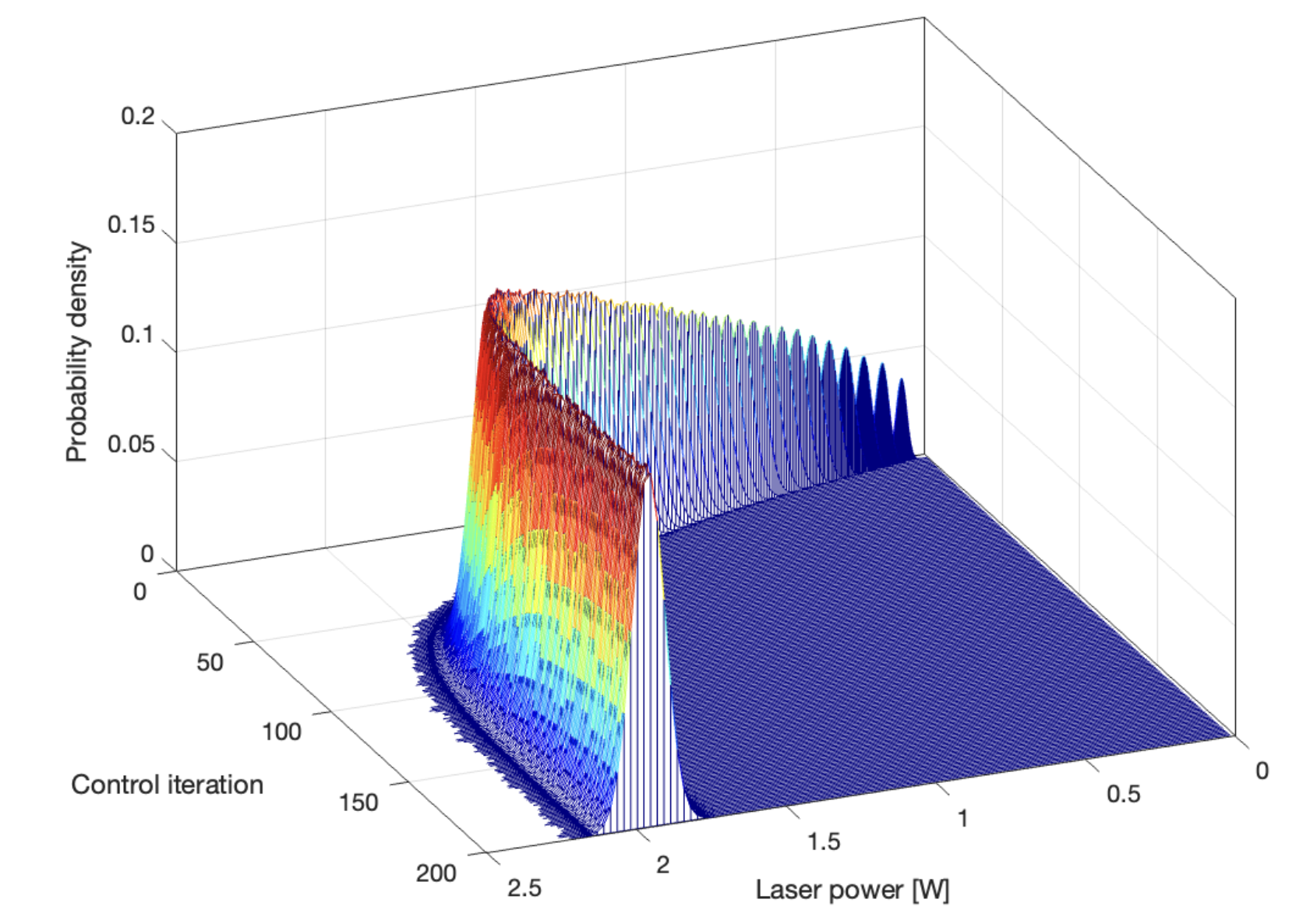}
        \caption{2D view of the laser power distributions at each control iteration in the presence of Gaussian measurement uncertainty.}
    \end{subfigure}
    \caption{The observed range of error fluctuations at each control iteration (top figure) reflects the dynamic response of the control system. As the control system progressively minimizes tracking error, the distributions in final control iterations exhibit a more constrained variability. The results show the sensitivity of the tracking error to the temperature measurement uncertainty in the temperatures over time. The historical evolution of laser power distributions (bottom figure) is highly influenced by the tracking error distributions. The progression of control iterations refines the laser power range from 0\,W  to 0.28\,W to a more focused distribution between approximately 1.8\,W and 2.3\,W.}~\label{fig:error-gaussian-2d-3d}
\end{figure}
\begin{figure}[!b]
    \centering
    \begin{subfigure}{0.9\linewidth}
        \centering
        \includegraphics[width=\linewidth, trim={0cm 0cm 0cm 0cm}, clip]{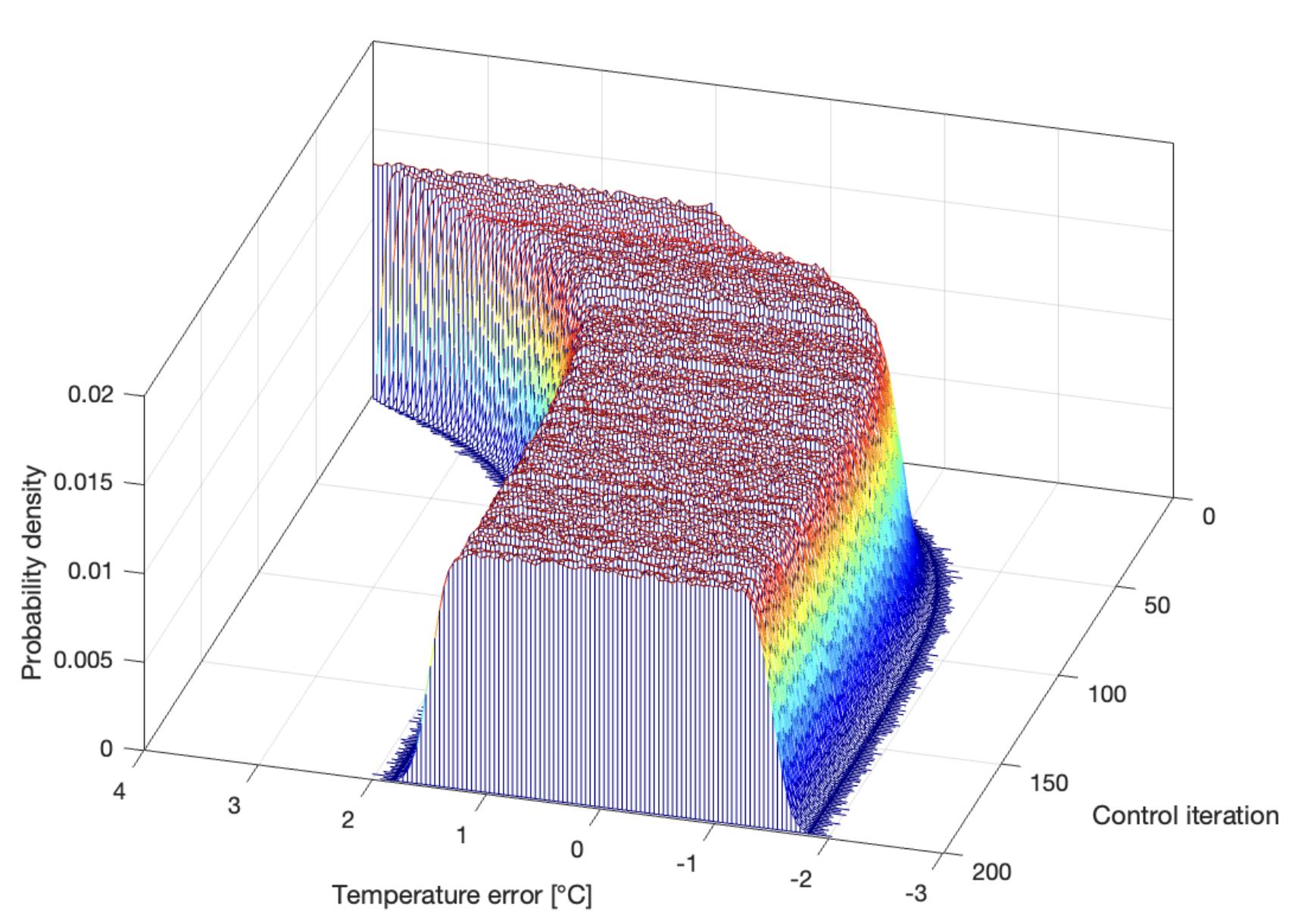}
        \caption{3D view of the tracking error distributions at each control iteration in the presence of uniform temperature measurement uncertainty.}
    \end{subfigure}
    \begin{subfigure}{0.95\linewidth}
        \centering
        \includegraphics[width=\linewidth, trim={0cm 0cm 0cm 0cm}, clip]{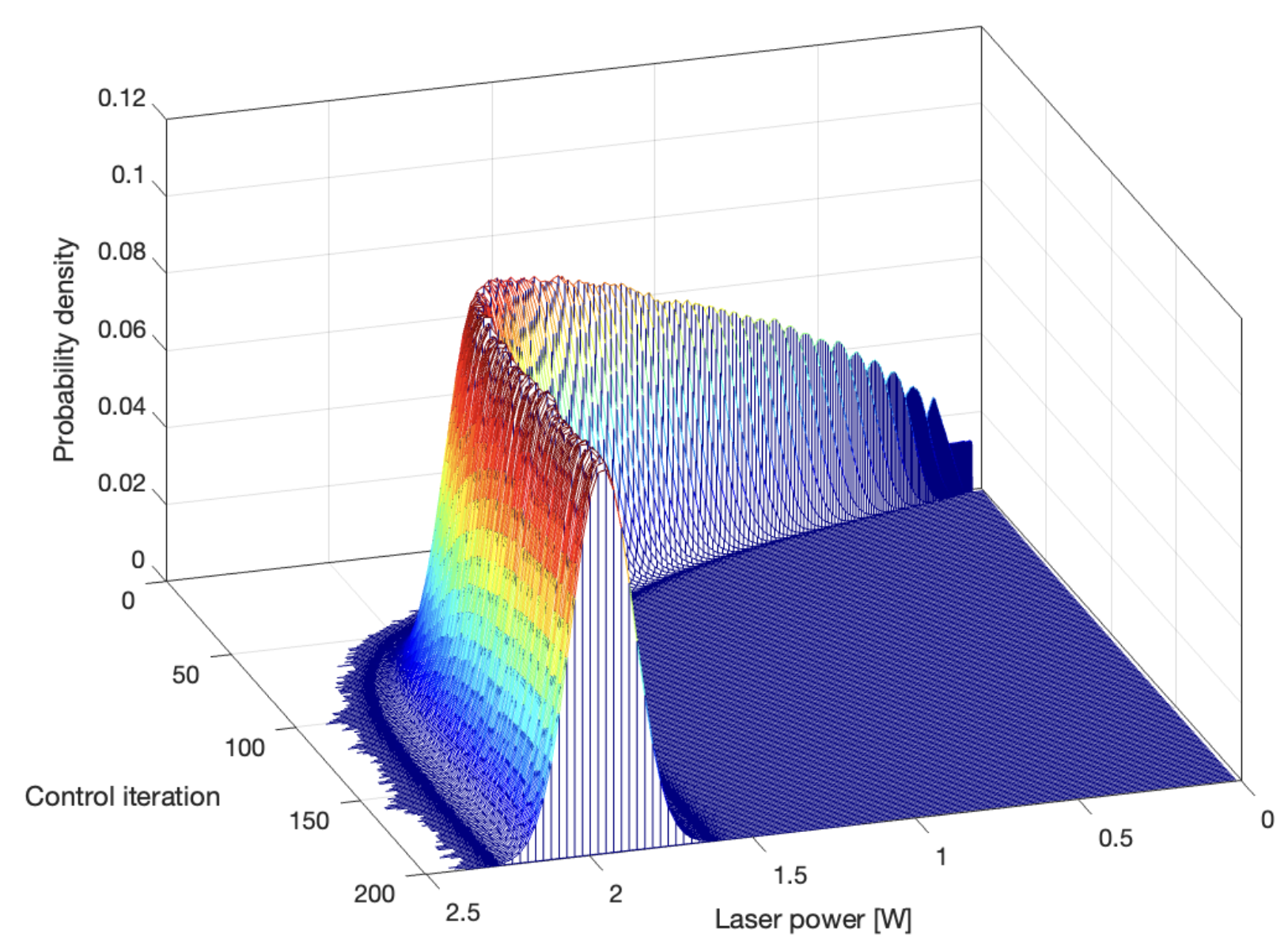}
        \caption{2D view of the laser power distributions at each control iteration in the presence of uniform temperature measurement uncertainty.}
    \end{subfigure}
    \caption{The distribution of the tracking error (top figure) at each control iteration reflects the sensitivity of the error to the uniform temperature measurement uncertainty in the measured temperature. As the control system gradually minimises the tracking error, the distributions in the final control iterations show a more restricted variability. The historical evolution of the laser power distributions (bottom figure) shows that the measurement uncertainty can strongly influence the laser power as output from the control system. The results show that in the early iterations, the laser power has large variations between zero and 2.3\,W, whereas in the last iterations, it ranges between 1.6\,W and 2.3\,W.}~\label{fig:laser-uniform-2d-3d}
\end{figure}

In Figure~\ref*{fig:error-gaussian-2d-3d}, we observe how the tracking error and laser power change over control iterations with Gaussian uncertainty in the measured temperatures. The results in Figure~\ref*{fig:error-gaussian-2d-3d} (a) indicate that at each control iteration, there's a noticeable range of error variation. 
As the control system minimises the tracking error, the distributions in the final control iterations have less variability. Initially, the main range of error variation goes from 0\,\textdegree C to 5\,\textdegree C, but it narrows to -2.5\,\textdegree C to 2.5\,\textdegree C in the final iterations. Changes in tracking error due to the temperature measurement uncertainty can influence the distribution of laser power, and the control system's output. In Figure~\ref*{fig:error-gaussian-2d-3d} (b), we observe the historical changes in the laser power distribution from the initial to the final control iteration. The variations start between zero and 0.3\,W, covering the lower and upper limits of laser power. The evolution concludes with a more defined range, approximately 1.8 W\,to 2.3\,W. 

The relationship between laser power and spot temperature in the SLS process is complex. Laser power directly controls the energy input into the powder bed, resulting in increased energy absorption and consequently increased temperatures at the laser spots. These temperatures play a direct role in material fusion and consolidation. Laser power impacts the distribution of spot temperatures, influencing material properties and the overall quality of the sintering process. The fluctuations in laser power determined by the control system, due to uncertainty in the measured temperature, suggest potential deviations from the optimal laser power. 

The lack of temperature uniformity, coupled with laser power fluctuations, may prevent effective powder bonding (at low laser powers) or result in overheating and defect formation (at high laser powers). This variability can impact the mechanical strength of printed objects.

Figure~\ref*{fig:laser-uniform-2d-3d} shows how tracking error and laser power change over control iterations when there is uniformly-distributed temperature measurement uncertainty. In Figure~\ref*{fig:laser-uniform-2d-3d} (a), as the control system minimizes tracking error, we observe noticeable reductions in the variability of the error distribution during the final control iterations. The main range of error variations initially spans from 1\,\textdegree C to 4\,\textdegree C, gradually narrowing to -2\,\textdegree C to 2\,\textdegree C in the last iterations. In Figure~\ref*{fig:laser-uniform-2d-3d} (b), we take a close look at how laser power changes over control iterations  when there is uniformly-distributed temperature measurement uncertainty. Starting from zero to 0.28\,W, the laser distribution varies over time. The data indicates that the laser fluctuates between around 1.6\,W and 2.3\,W in the last iterations. 

Compared to the Gaussian temperature measurement uncertainty, the tracking error exhibits a narrower range of variations at final iterations, suggesting smoother temperature distributions across different laser spots. However, there are larger fluctuations in laser power compared to those observed under Gaussian uncertainty. Similarly, these fluctuations in both error and laser power due to uniform temperature measurement uncertainty can affect the quality of printed objects by inducing underheating or overheating during the sintering process.

% Grammar:
% \input{appendix.tex}

\end{document}